\documentclass[%
reprint,
twocolumn,
aps,
pre,
showpacs,
notitlepage,
nobibnotes,
superscriptaddress
]{revtex4-2}

\usepackage{float}
\usepackage{bigints}
\usepackage{psfrag}
\usepackage{grffile}
\usepackage{verbatim}
\usepackage{microtype}
\usepackage{multirow}
\usepackage{enumitem}
\usepackage{amsmath}
\usepackage{amssymb}
\usepackage{amsthm}
\usepackage{mathrsfs}
\usepackage{mathtools}
\usepackage{graphicx}
\usepackage{bbm}
\usepackage{subfig}
\usepackage[linesnumbered,ruled]{algorithm2e}
\usepackage{color}
\usepackage{tcolorbox}

\usepackage{ragged2e}

\definecolor{myblue}{rgb}{0.153,0.322,0.706}
\usepackage[colorlinks,linkcolor=myblue,urlcolor=myblue,citecolor=myblue]{hyperref}
\usepackage{geometry}
\usepackage{url}
\usepackage{hyperref}
\usepackage{xcolor}

\newcommand{\be}{\begin{equation}}
\newcommand{\ee}{\end{equation}}

\def\bc{\begin{center}}
\def\ec{\end{center}}
\def\bea{\begin{eqnarray}}
\def\eea{\end{eqnarray}}

\DeclareMathOperator{\arcsinh}{arcsinh}
\DeclareCaptionJustification{justified}{\justifying}
\definecolor{airforceblue}{rgb}{0.36, 0.54, 0.66}
\definecolor{brickred}{rgb}{0.8, 0.25, 0.33}
\definecolor{amber}{rgb}{1.0, 0.75, 0.0}
\definecolor{applegreen}{rgb}{0.55, 0.71, 0.0}
\definecolor{magenta}{rgb}{0.965, 0, 0.859}

\newif\ifshowcomments
\showcommentstrue  
\ifshowcomments
    
    \newcommand{\jpg}[1]{\textcolor{blue}{\bf [JPG: #1]}}        
\else
    \newcommand{\jpg}[1]{}
    \newcommand{\eg}[1]{}
    \newcommand{\jfm}[1]{}
\fi

\newcommand{\newtext}[1]{{\color{black}{#1}}}
\newcommand{\newnewtext}[1]{{\color{black}{#1}}}
\newcommand{\deletetext}[1]{}

\newcommand{\E}{{\mathcal E}}
\newcommand{\ssp}{\nobreak\hspace{0.1em}}
\newcommand{\er}[1]{Eq.\ssp\eqref{#1}}
\newcommand{\ers}[2]{Eqs.\ssp(\ref{#1}-\ref{#2})}

\begin{document}

\title{Level 2.5 large deviations and uncertainty relations for self-interacting jump processes: tilting constructions and the emergence of time-scale separation}

\author{Francesco Coghi}
\email{francesco.coghi@leicester.ac.uk}
\affiliation{School of Physics and Astronomy, University of Nottingham, Nottingham, NG7 2RD, UK}
\affiliation{Centre for the Mathematics and Theoretical Physics of Quantum Non-Equilibrium Systems,
University of Nottingham, Nottingham, NG7 2RD, UK}

\author{Juan P.\ Garrahan}
\affiliation{School of Physics and Astronomy, University of Nottingham, Nottingham, NG7 2RD, UK}
\affiliation{Centre for the Mathematics and Theoretical Physics of Quantum Non-Equilibrium Systems,
University of Nottingham, Nottingham, NG7 2RD, UK}

\date{\today}

\begin{abstract}
Self-interacting jump processes (SIJPs) describe systems with non-Markovian stochastic dynamics in which transition rates depend on empirical observables of the process, which gives rise to long-range memory and feedback. We derive the ``level-2.5'' large deviation (LD) principle governing the joint fluctuations of empirical occupation measure and the flux matrix for a broad class of SIJPs with general functional dependence on an empirical observable. The derivation is based on an exponential tilting construction and reveals a separation between a faster timescale of the microscopic dynamics and a slower timescale of the memory-driven evolution of transition rates, which is expressed through an exponentially discounted LD rate functional. Using this variational framework, we derive kinetic and thermodynamic uncertainty relations that extend classical Markovian bounds to non-Markovian systems, and illustrate their performance with simple examples. 
\end{abstract}

\maketitle

This is a companion paper to Ref.~\cite{coghi2026level}. Here we present the full derivations of the results reported in an abridged manner in Ref.~\cite{coghi2026level}, together with additional general results and illustrative examples, in a self-contained form.

\bigskip

\section{Introduction}
\label{sec:intro}

Communicating information about past experiences is a distinctive feature of intelligent systems. While many organisms store and process information internally via a nervous system, numerous others are too small to possess such structures. Over evolutionary timescales, these organisms have instead developed forms of \textit{external spatial memory}, whereby information is encoded in the surrounding environment rather than stored internally. Several micro-organisms, including \textit{E.\ Coli}, \textit{P.\ Aeruginosa}, and \textit{Pharaoh's ants}, achieve this by depositing chemical traces~\cite{budrene1995dynamics,brenner1998physical,sumpter2003from,jackson2006longevity,zhao2013psl-trails,gelimson2016multicellular}, while others, such as slime moulds~\cite{reid2012slime}, encode information through persistent structural modifications of their environment. These environmental changes act as cues that can be sensed either by conspecifics or by the same organism at later times. In the latter case, the resulting dynamics are described as \textit{autochemotactic} or \textit{self-interacting}, whereby an organism’s future behaviour is influenced by its interaction with an environment it has itself modified. Remarkably, analogous behaviours have recently been realised in artificial systems, including autonomous agents such as self-propelled particles~\cite{howse2007self-motile,thutupalli2011swarming,hokmabad2022chemotactic,nakayama2023tunable} and self-morphic active matter~\cite{kumar2024emergent}.

Approximate models of the dynamics of self-interacting autochemotactic agents exist. Reinforced random walks~\cite{coppersmith1986random,toth2001self-interacting,othmer2006aggregation,pemantle2007a-survey,erschler2011stuck,kious2016stuck,barbier-chebbah2022self-interacting,bremont2024exact}, both on edges and vertices, and field-theoretic descriptions~\cite{keller1970initiation,tsori2004self-trapping,gelimson2015collective,grafke2017spatiotemporal}, also in combination with Langevin-type dynamics~\cite{grima2005strong-coupling,sengupta2009dynamics,pohl2014dynamic,kranz2016effective} can replicate the long-time behaviour of self-interacting agents, which reveal rich dynamical features. These models are extremely valuable in mimicking self-attracting and self-repelling behaviour and in replicating the formation of specific trails and patterns. A rigorous description is offered by self-interacting processes (SIPs, see~\cite{pemantle2007a-survey} for an introduction), a class of reinforced processes, such as chains~\cite{moral2007self-interacting,kannan2008self-interacting,coghi2025self-interacting}, jump processes~\cite{budhiraja2025jump}, and diffusion~\cite{benaim2002self-interacting}, whereby the self-interaction is mediated by a functional of the \textit{empirical occupation measure}, a random field that models the chemical, or structural, changes generated by autochemotactic agents.

Methods such as stochastic approximation~\cite{benaim1999dynamics,benaim2002self-interacting} and hydrodynamic limits are often effective in studying the asymptotic behaviour of the models above, including stationary dynamics and typical fluctuations (see, e.g.,~\cite{benaim2002self-interacting,kurtzmann2010the-ode-method,benaim2011self-interacting, coghi2024current}). By contrast, finite-time fluctuations and rare events have only recently begun to attract attention~\cite{franchini2017large,budhiraja2022empirical,coghi2024current,coghi2025self-interacting,budhiraja2025large,coghi2025accelerated,budhiraja2025jump,coghi2026level}, despite their known importance in shaping future dynamics. This difficulty arises because SIPs, as well as other reinforced random walks and related field-theoretic models, are inherently non-Markovian when restricted to the configuration space accessible to the agent. This structure often obstructs straightforward analytical treatments and gives rise to complex phenomena, such as dynamical ergodicity breaking~\cite{benaim2002self-interacting,benaim2005self-interacting,coghi2024current} and modified first-passage dynamics~\cite{aleksian2024self-interacting,coghi2025accelerated}.

To address this gap, here we derive the large deviation asymptotics of the empirical occupation and flux, that is, the dynamical large deviations (LDs) at ``level-2.5'' (see Refs.~\cite{maes2008canonical,bertini2012large} for the Markovian case), for a broad class of self-interacting jump processes (SIJPs), in which the self-interaction depends functionally on both the empirical occupation measure and the empirical flux. (Our main results here were presented in shortened form, and without detailed proofs, in our recent Ref.~\cite{coghi2026level}.)

Our approach is based on a formal tilting procedure and, under suitable assumptions, clarifies the intrinsic multi-scale dynamics characteristic of these systems. We then study generalised scalar flow- and current-like observables and derive variational bounds on their fluctuations, yielding SIJP generalisations of kinetic and thermodynamic uncertainty relations~\cite{barato2015thermodynamic,gingrich2016dissipation,garrahan2017simple,terlizzi2019kinetic,horowitz2020thermodynamic,macieszczak2024ultimate,prech2024optimal}. We thus extend the fluctuation-theoretic tools developed for Markov processes to non-Markovian dynamics. Finally, through representative examples, we show how self-interaction can enhance fluctuations relative to purely Markovian systems, and assess how exact or numerical large deviation results compare with the SIJP fluctuation bounds.

The paper is structured as follows. In Sec.~\ref{sec:Back} we give a non-exhaustive overview of the statistical physics literature on fluctuations of non-Markov processes, in Sec.~\ref{sec:model} we introduce the class of non-Markov SIJPs we focus on, and in Sec.~\ref{sec:25} we derive their level-2.5 LDs. By restricting to specific classes of SIJPs, in Sec.~\ref{sec:Kin} we derive the probability bounds for flux observables: specifically, in Subsec.~\ref{subsec:KUR} we derive a SIJP version of the kinetic uncertainty relation (KUR) \cite{garrahan2017simple,terlizzi2018kinetic} which we call SIJP-KUR, and in Subsec.~\ref{subsec:UKUR} a SIJP version of the ``ultimate KUR'' (UKUR) \cite{macieszczak2024ultimate,prech2024optimal} which we call SIJP-UKUR. We then look at the case of generalised current observables in Sec.~\ref{sec:Thermo} and derive a SIJP version of the thermodynamic uncertainty relation (TUR) \cite{barato2015thermodynamic,gingrich2016dissipation}, which we call SIJP-TUR. In Sec.~\ref{sec:Ex} we illustrate our results with simple examples. Finally, in Sec.~\ref{sec:Conclusion} we conclude the paper with a summary and an overview of interesting research directions. 

\bigskip

\noindent
\textbf{List of acronyms:} 
FDT~=~fluctuation-dissipation theorem.
GLE~=~generalised Langevin equation.
KUR~=~kinetic uncertainty relation.
LD~=~large deviation.
SIJP~=~self-interacting jump process.
SIP~=~self-interacting process.
TUR~=~theromdynamic uncertainty relation.
UKUR~=~ultimate kinetic uncertainty relation.

\section{Related work on fluctuations of non-Markov processes}
\label{sec:Back}

Long-range memory effects, characteristic of non-Markovian models, are known to strongly influence fluctuation behaviour, leading to departures from the large deviation theory established for Markovian systems~\cite{lecomte2007thermodynamic,garrahan2007dynamical,maes2008canonical,touchette2009the-large,bertini2012large,chetrite2015nonequilibrium,chetrite2015variational,garrahan2018aspects,barato2018a-unifying,jack2020ergodicity,carugno2022graph-combinatorial}. In SIPs (and SIJPs), such memory effects originate from the explicit dependence of the dynamics on the empirical occupation measure (and empirical flux).

Non-Markovian random walks are known to exhibit memory-induced phenomena~\cite{schutz2004elephants,rebenshtok2007distribution,harris2015random,jack2020giant}, and a ``temporal additivity principle'' has been proposed to partially account for fluctuations of time-additive observables~\cite{harris2009current}. These observables—including the empirical occupation measure—were argued to become slow variables with vanishing noise in the long-time limit, leading to reduced fluctuations and reinforced dynamics that progressively constrain exploration. Within this framework, mechanisms generating rare events in non-Markovian random walks, such as the elephant random walk and its Gaussian variant~\cite{jack2020giant}, have been identified.

Despite these advances, a comprehensive understanding of fluctuation dynamics in SIPs---where the evolution intrinsically depends on slowly varying variables---remained elusive until very recently. Large-deviation principles for self-interacting Markov chains~\cite{budhiraja2022empirical,budhiraja2025large} and jump processes~\cite{budhiraja2025jump,coghi2026level}
have now been established. In particular, Ref.~\cite{budhiraja2025jump} rigorously proves a level-2.5 large deviation principle for SIJPs under a self-interaction affine in the empirical occupation measure. In Ref.~\cite{coghi2026level} we extend this result to general functionals of empirical occupation measure and flux, showing that fluctuations of a SIJP are \textit{optimally} sampled by a non-homogeneous Markov process. Together with related large deviation results for self-interacting Markov chains~\cite{budhiraja2022empirical,budhiraja2025large}, generalised Pólya urn models~\cite{franchini2017large,franchini2025corrigendum}, non-homogeneous (periodically-driven) Markov processes~\cite{barato2018current}, and semi-Markov processes~\cite{maes2009dynamical,garrahan2010thermodynamics,shreshtha2019thermodynamic,carollo2019unraveling,ertel2022operationally,jia2022large,macieszczak2024ultimate,liu2024semi-markov,maier2025a-pedestrians}, these works provide the foundations for a coherent framework to study fluctuations in non-Markovian systems.

The loss of Markovianity introduces substantial technical challenges in the analysis of fluctuations. Alongside large deviation theory, fluctuation–dissipation theorems and uncertainty relations play a central role, as they provide general principles linking fundamental macroscopic observables to their typical fluctuations. As in the Markov case~\cite{barato2015thermodynamic,gingrich2016dissipation,garrahan2017simple,terlizzi2019kinetic,horowitz2020thermodynamic,macieszczak2024ultimate}, these results can often be derived variationally from level-2.5 LDs. Despite sustained theoretical interest, progress in this direction remains limited for non-Markovian dynamics, with most existing results confined to generalised Langevin equations (GLEs) and semi-Markov processes.

Work on FDTs in non-Markov systems has mostly proceeded by embedding memory into GLEs or semi-Markov descriptions and then asking which parts of the Markovian FDT survive. For diffusions with memory kernels, it was shown that memory generally enhances the \textit{frenetic}, or time-symmetric dynamical activity, of the linear response~\cite{maes2013fluctuation-response}. For general time dependence, the \textit{second} FDT associated to GLEs can be derived via Mori--Zwanzig projections~\cite{mori1965transport,zwanzig1973nonlinear,espanol2009mori--swanzig}, with notable limitations when coarse-graining is in place~\cite{schilling2022coarse-grained,jung2022non-markovian,zhu2023general}, even in the case non-equilibrium baths are introduced~\cite{maes2014on-the-second}. Beyond diffusions, related fluctuation–response structures have been derived for semi-Markov networks~\cite{chen2022generalized}.

For uncertainty relations, the classical TUR~\cite{barato2015thermodynamic,gingrich2016dissipation} for Markov jump and diffusion processes has been generalised along three main non-Markov directions: delay, semi-Markov kernels, and quantum/non-Markovian baths. For time-delayed Langevin systems an uncertainty relation was derived by bounding fluctuations in terms of the Kullback--Leibler divergence between a path measure and its time-reversed counterpart~\cite{vu2019uncertainty,hasegawa2019fluctuation,rosinberg2018comment}. For antisymmetric observables this yields a bound that reduces to the standard TUR in the Markov limit. For semi-Markov jump processes, TURs that relate entropy production to the current asymptotic variance were derived using renewal theory~\cite{shreshtha2019thermodynamic} and an alternative derivation~\cite{ertel2022operationally} that does not use the time-direction independence~\cite{chari1994on-reversible,maes2009dynamical}. 

\newtext{Notably, in deriving TURs, it is widely recognised that non-Markovian dynamics complicates the definition of entropy production, since irreversibility cannot in general be assigned from an instantaneous state alone. Nevertheless, several works have shown that local or instantaneous entropy-production rates can still be obtained once the memory structure is properly accounted for. This is the case, for example, for active Ornstein--Uhlenbeck particles~\cite{Caprini2019} and GLE dynamics with memory~\cite{di-terlizzi2020a-thermodynamic}. In both cases, irreversibility is quantified through a comparison of path weights, or equivalently through a dynamical cost associated with an auxiliary representation of the process.}

For the case of KURs in the same semi-Markov and renewal settings, they have been obtained bounding current fluctuations in terms of effective activities depending on the full waiting-time statistics rather than just mean jump counts~\cite{garrahan2017simple,terlizzi2019kinetic} or mean waiting times~\cite{macieszczak2024ultimate}, providing natural non-Markov analogues of Markovian KURs~\cite{shreshtha2019thermodynamic,ertel2022operationally}. Related kinetic-type bounds also appear in delayed and quantum transport models, where fluctuation-theorem or full-counting-statistics approaches yield constraints on current Fano factors with explicit dependence on delay kernels or bath correlation functions~\cite{agarwalla2018assessing,vu2019uncertainty,hasegawa2019fluctuation}. Additionally, stricter TURs that hold if the observed semi-Markov process arises from coarse-graining an underlying Markov chain have also been established, and their violation can be used as diagnostic of genuine memory~\cite{ertel2022operationally}.

Finally, in Ref.~\cite{coghi2026level}, we extended this body of work by deriving uncertainty relations for general SIJPs. In the following, we present the detailed derivations underlying the SIJP-KUR and SIJP-TUR reported in~\cite{coghi2026level}, and introduce the SIJP-UKUR together with its derivation.

\section{Self-interacting jump processes}
\label{sec:model}

We define a SIJP as the continuous-time chain $\left( X_t \right)_{0 \leq t \leq T}$ of discrete configurations, $ x \in \mathcal{S} \coloneqq \left\lbrace 1,2,\cdots, d \right\rbrace$, with $\E \coloneqq \left\lbrace (x,y) \in \mathcal{S} \times \mathcal{S}: x \neq y \right\rbrace $ denoting the set of all possible jumps between them. At time $t$, the SIJP takes a jump from the current state $x$ to $y \neq x$ with evolving rate matrix (stochastic generator) $Q_{xy}(A_t) \in \mathbb{R}_+$, that depends on a general state dependent \textit{empirical observable} of the process
\begin{equation}
    \label{eq:EmpDepSIJP}
    A_t = 
    t^{-1} \int_0^t 
        f_{X_{t'}} \, dt' 
    + 
    t^{-1} \sum_{\substack{0 \leq t' \leq t \\ (X_{t'_-}, X_{t'}) \in \E}} 
        g_{X_{t'_-}X_{t'}} \, ,  
\end{equation}
\newtext{where $t'$ are jump times, and}\deletetext{with} $f: \mathcal{S} \rightarrow \mathbb{R}$ and $g: \E \rightarrow \mathbb{R}$ bounded functions. \newtext{We assume the mapping $Q(a)$ to be at least of class $C^1$ on $\mathbb{R}$ as this guarantees enough regularity for the existence of a stationary state. Furthermore, for the block-averaging argument that comes below, we assume $Q_{xy}(a)$ to be uniformly positive on the admissible domain of $a$. This means
\begin{equation}
    Q_{xy}(a) \geq q_- > 0 \, ,
\end{equation}
for all $(x,y) \in \mathcal{E}$ and relevant $a$, with $q_-$ some fixed constant independent of $a$. This means that irrespective of the argument of $Q(\cdot)$, the transition probabilities for the directed links in $\E$ are never vanishing and always at least $q_-$. If $g=0$ in Eq.\ \eqref{eq:EmpDepSIJP}, such a condition simply follows from continuity of the map $Q(a)$ as well as $f$ being bounded, otherwise it must be imposed.}

By conservation of probability, the diagonal terms of the rate matrix are the negative escape rates 
\begin{equation}
    \label{eq:DiagRateSIJP}
    Q_{xx}(A_t) = - \sum_{y \in \mathcal{S}, y \neq x} Q_{xy}(A_t) \, .
\end{equation}
In contrast with a homogeneous Markov process, the jump rates are not constant over time and they change as $A_t$ evolves, making the process non-Markovian.

We define
\begin{equation}
    \label{eq:EmpOccSIJP}
    L_x(t) = t^{-1} \int_0^t \mathbf{1}_x(X_t') \, dt' \, ,
\end{equation}
as the \textit{empirical occupation measure} of the process, viz.\ the fraction of time that the process has spent at each and every configuration of the state space up to time $t$, with $\mathbf{1}_x(X_{t'})$ the indicator function of configuration $x$. Such a time-evolving object lives in a subset of the whole probability space on $\mathcal{S}$, i.e., $L(t) \in \mathcal{P}_0(\mathcal{S}) \subseteq \mathcal{P}(\mathcal{S})$. Along with it, we introduce the empirical flux
\begin{equation}
    \label{eq:EmpFluxSIJP}
    \Phi_{xy}(t) = t^{-1} \sum_{\substack{0 \le t' \le t \\ 
(X_{t'_-}, X_{t'}) \in \E} } \mathbf{1}_{x}(X_{t_-'}) \mathbf{1}_{y}(X_{t'}) \, ,
\end{equation}
viz.\  the non-negative $d \times d$ matrix (with zero diagonal elements) of time-averaged number of jumps in the interval $[0,t]$ between any two different configurations in $\mathcal{S}$.

Given the two objects above, it is possible to cast $A_t$ in \er{eq:EmpDepSIJP} as a function of them as follows:
\begin{equation}
    \label{eq:EmpDepSIJPConfig}
    A_t = \sum_{x \in \mathcal{S}} f_x L_x(t) + \sum_{(x,y) \in \E} g_{xy} \Phi_{xy}(t) \, .
\end{equation}
Such a rewriting highlights the dependence on the specific stochastic features of the SIJP that we consider in this work. As usual, if one moves to an extended space where the state is $\left( X_t, L(t), \Phi(t) \right)_{0 \leq t \leq T}$, then a Markovian description of the dynamics can be constructed recursively from one jump to the next by a standard procedure.

In the main text of Ref.~\cite{coghi2026level}, the observable $A_t$ is restricted to the case $g_{xy}=0$ for all $(x,y) \in \E$ (and only at the end we briefly mention the extension to the case with a non-zero $g$ dependence). Here, we consider from the outset the most general class of SIJPs as the derivation of the level-2.5 LDs appearing in Sec.~\ref{sec:25} follows the same lines.

To proceed with the derivation, we need to make a few \newtext{other} assumptions. First, we assume that, for every value $a$ in the range relevant to the observable, the rate matrix $Q(a)$ is irreducible---and therefore positively recurrent, since $\mathcal{S}$ is finite. \deletetext{Then, we assume}\newtext{For each frozen value of $a$, this guarantees the existence of a unique \textit{accompanying distribution} $\gamma(a)$ satisfying
\begin{equation}
    \gamma(a) Q(a)=0,
    \qquad 
    \sum_{x\in\mathcal S}\gamma_x(a)=1.
\end{equation}
The $C^1$ regularity of the map $a\mapsto Q(a)$ then implies that $a\mapsto \gamma(a)$ is continuous. Although irreducibility and regularity together imply the existence of at least a limiting stationary state via Brouwer's fixed-point theorem, they do not imply that the self-interacting dynamics converges. We therefore need to further assume that} both the empirical measure and flux converge in the long-time limit to \newtext{finite} typical values, i.e.,
\begin{equation}
    \label{eq:EmpTypValues}
    L(t) \rightarrow \pi \hspace{1cm} \Phi(t) \rightarrow \varphi \, ,
\end{equation}
for some pair $(\pi,\varphi)$, which may depend on the trajectory and is not necessarily unique. \newtext{These assumptions can be checked on a case-by-case basis and in Appendix~\ref{app:ass} we discuss them more in detail and provide simple examples in which they fail.}

We remark that for a Markov process with fixed generator $Q$, the stationary state $\pi$ obeys $\pi Q = 0$, the empirical measure converges to $\pi$, and the empirical flux to $\pi \circ Q$ (where $\circ$ denotes the Hadamard product). In the SIJP case things are different: one needs the long-time limit of the observable, i.e.,
\begin{equation}
    \label{eq:LongTimeObs}
    A_t \rightarrow \bar{a} = \sum_{x \in \mathcal{S}} f_x \pi_x + \sum_{(x,y) \in \E} g_{xy} \varphi_{xy} \, ,
\end{equation}
and the stationary density $\pi$ solution of the nonlinear self-consistency equation
\begin{equation}
    \label{eq:StatSIJP}
    \pi Q(\bar{a}) = 0 \, .
\end{equation}
Also, $\varphi \neq \pi \circ Q(\bar{a})$ in general; it will be given an exact form in the next Section\deletetext{when restricting to a unique fixed point}. \newtext{In particular, the limiting empirical flux has to satisfy the stationarity constraint
\begin{equation}
    \label{eq:FluxConservation}
    \sum_{y\neq x}\varphi_{xy}
    =
    \sum_{y\neq x}\varphi_{yx},
    \qquad x\in\mathcal S,
\end{equation}
but it need not be fixed a priori by the typical Markov form $\pi_x Q_{xy}(\bar a)$ when considering the more general stationary pairs appearing in the level-2.5 variational problem that follows.}

In the following, we denote by $\mathcal{A}_{\mathrm{stat}}$ the set of stationary limiting pairs $(\pi,\varphi)$ satisfying \eqref{eq:LongTimeObs}, \eqref{eq:StatSIJP}, and \eqref{eq:FluxConservation}. We also highlight that, differently from the more usual time-homogeneous case, the selection of the specific stationary pair $(\pi,\varphi)$ is not ruled by a breaking of the irreducibility of the process over the state space---we assume above that irreducibility always holds---but it is a dynamical feature emerging from the interplay of noise-induced fluctuations and long-range memory.

\subsection{Notation}

The notation we use is as follow. We use upper case for random variables (or functions and sets) and lower case for their realisations, while reserving $P$ for probability. \newtext{$\mathcal{P}$ is used to refer to a probability space and, generally, $P$ for probability distributions, with $\mathbb{P}$ specially reserved for path probability measures.} We use a compact notation for multidimensional objects: given vector $\pi \coloneqq (\pi_x)_{x \in \mathcal{S}}$ and matrix $Q \coloneqq (Q_{xy})_{(x,y) \in \mathcal{S} \times \mathcal{S}}$, $\pi Q$ denotes their internal product, $(\pi Q)_y =
\sum_{x \in \mathcal{S}} \pi_x Q_{xy}$, and use $\circ$ for the Hadamard product, with $ \pi \circ Q$ having elements $\pi_x Q_{xy}$. For scalar quantities we write the time dependence with a subscript, e.g., $A_t$, for compactness, whereas for higher-dimensional obejcts, say $\mu \coloneqq (\mu_x)_{x \in \mathcal{S}}$ as $\mu(t)$. $\mathbf{1}_x(\cdot)$ is the indicator function, which gives one if the argument coincides with $x$ or zero otherwise. We use overbar to mark typical values, e.g., $\lim_{t \rightarrow \infty} A_t = \bar{a}$. The notation $a_T \sim b_T$ as $T \to \infty$ means asymptotic equivalence, namely $\frac{a_T}{b_T} \to 1$. In particular, $P_T(a) \sim 1$ means that $P_T(a) \to 1$ as $T \to \infty$. Similarly, the symbol $\asymp$ indicates equality up to sub-exponential factors in a large parameter (often time), so that $P_T(\cdot) \asymp e^{- T I(\cdot)}$ is
equivalent to $I(\cdot) = -\lim_{T \rightarrow \infty} T^{-1} \log P_T(\cdot)$. We define the asymptotic variance (or susceptibility) of an (intensive) trajectory observable such as $B_T$ as $\text{var}(b) \coloneqq \lim_{T \rightarrow \infty} T \, \text{var}(B_T)$ where $\text{var}(B_T) = \mathbb{E} \left[ (B_T - \bar{b})^2 \right]$. Finally, we use the row convention: probabilities are row vectors, stochastic generators have rows adding up to zero, and time-propagation is left to right,
e.g., $\partial_t \mu(t) = \mu(t) Q$.]

\section{Level 2.5 large deviations: slow/fast dynamics} 
\label{sec:25}

\subsection{Probability path measures and tilting}


Our focus is on the long, but finite, time dynamics of the empirical occupation measure in \er{eq:EmpOccSIJP} and the empirical flux in \er{eq:EmpFluxSIJP}. By going beyond their typical behaviour, we formally derive the asymptotic exponential scaling of the joint probability $P(L(t) = \ell, \Phi_t = \phi) \eqqcolon P_t(\ell,\phi)$, namely the level-2.5 large deviation rate function for SIJPs. 


To do so, we use exponential tilting, a well-known technique in statistical physics. To our knowledge, this method has hitherto been applied to derive the level-2.5 LDs of Markov processes only, including jump processes~\cite{de2001large,maes2008canonical,bertini2012large,barato2015a-formal,chetrite2015variational} and diffusions~\cite{kusuoka2010large,barato2015a-formal,chetrite2015variational}. 
[Note that for discrete-time Markov chains\newtext{, to study the statistics of $k$-th order empirical occupation measures, such as the ($k=2$) number of jumps between any two states,} it is enough to \textit{magnify} the state space\newtext{, viz.\ replace original states with states obtained by $k$ steps in the chain} \deletetext{to study large deviations}]. In what follows, by properly modifying the assumptions in the derivation, we will extend the use of tilting to non-Markov jump processes.


With the definitions above we can write the path probability measure: given a path $\omega = \left\lbrace (x_0,\deletetext{\tau_0}\newtext{t'_0}),(x_1,\deletetext{\tau_1}\newtext{t'_1}), \cdots, (x_n, T) \right\rbrace$ that is a specific realisation of the SIJP $\left( X_t \right)_{0 \leq t \leq T}$, where \deletetext{$\tau_{i-1}$}\newtext{$t'_{i-1}$} is the $i$-th jump time, the corresponding path measure reads
\begin{equation}
    \label{eq:PathProbaSIJP}
    \begin{split}
    \mathbb{P}_{T}(\omega) &= \\
    &\hspace{-1cm}\mu_0(x_0) \exp \left( \int_0^{\deletetext{\tau_0}\newtext{t'_0}} Q_{x_0 x_0} (A_s) \, ds \right) Q_{x_0 x_1}(A_{\deletetext{\tau_0}{\newtext{t'_0}}}) d\deletetext{\tau_0}\newtext{t'_0} \times \\ 
    &\hspace{0.8cm}\cdots \times \exp \left( \int_{\deletetext{\tau_{n-1}}\newtext{t'_0}}^{T} Q_{x_{n} x_{n}} (A_s) \, ds \right) \, ,
    \end{split}
\end{equation}
with $\mu_0$ the probability distribution of the initial state. Notice that the SIJP spends an exponentially distributed time at each state in $\mathcal{S}$. However, differently from the time-homogeneous case, the overall jump rate out of a state is random for a SIJP as rates change with time. As detailed in~\cite{daley2003an-introduction}, the overall jump rate out of $x$ is
\begin{equation}
    \label{eq:JumpOutRateSIJP}
    \Lambda_x(t,t+u)\coloneqq -\int_t^{t+u} Q_{xx}(A_s) \, ds \, ,
\end{equation}
where $t$ is the time at which the SIJP jumps onto $x$ and $u$ is the random time at which it jumps out of it. (In Appendix \ref{app:sim}, we describe two ways to simulate a SIJP.)

Thanks to the stationarity condition \eqref{eq:StatSIJP}, the law of large numbers implies that, in the long-time limit, the empirical measure and flux converge with probability tending to one to the set of stationary limiting pairs. Accordingly, rather than concentrating on a single pair $(\pi,\varphi)$, one has
\begin{equation}
    \label{eq:ProbaTypSIJP}
    P \!\left( (L(T),\Phi(T)) \in \mathcal{A}_{\mathrm{stat}} \right)
    \overset{T \rightarrow \infty}{\sim} 1 \, ,
\end{equation}
where the probability density above is a marginal of the path measure $\mathbb{P}_T$ in Eq.\ \eqref{eq:PathProbaSIJP}, and the dependence on the trajectory is implicit in the definitions of $L(t)$ and $\Phi(t)$ in Eqs.\ \eqref{eq:EmpOccSIJP} and \eqref{eq:EmpFluxSIJP}, respectively.


Along with the SIJP $\left( X_t \right)_{0 \leq t \leq T}$ we assume that a time-inhomogeneous Markov process $( \tilde{X}_t )_{0 \leq t \leq T}$ with path probability measure $\tilde{\mathbb{P}}_T$ and stationary state $\ell \neq \pi$ exists, such that
\begin{equation}
    \begin{split}
        \tilde{P}_T(\ell,\phi)
        &= \int \mathbf{1}_{\ell}(L(T)) \mathbf{1}_{\phi}(\Phi(T)) \, d \mathbb{\tilde{P}}_{T}(\omega)\overset{T \rightarrow\infty}{\sim} 1 \, ,
    \end{split}
    \end{equation}
for $\phi \neq \varphi$, in probability, and $0$ otherwise. We assume the dynamics of such a process to be generated by a time non-homogeneous rate matrix $\tilde{H}(t) \coloneqq ( \tilde{H}_{xy}(t) )_{(x,y)} \in \mathcal{S} \times \mathcal{S}$ with diagonal elements the negative escape rates
\begin{equation}
    \label{eq:DiagRateNonHomo}
    \tilde{H}_{xx}(t) = - \sum_{y \in \mathcal{S}: y \neq x} \tilde{H}_{xy}(t) \, ,
\end{equation}
\newtext{as well as uniformly positive jump rates, viz.\ 
\begin{equation}
    \tilde{H}_{xy}(t) \geq h_- > 0 \, ,
\end{equation}
for all $(x,y) \in \mathcal{E}$ and $t > 0$, with $h_-$ some fixed constant independent of $t$.}
Additionally, we assume that this time-inhomogeneous Markov process evolves on \deletetext{the same macroscopic} time-scale \newtext{comparable or slower than the time scale} of the original SIJP (\newtext{this requirement comes from the block-averaging derivation below and we postpone the detailed explained to then}\deletetext{as we explain in detail below}), and that an accompanying distribution $\tilde{\rho}(t)$\deletetext{, namely, a time-dependent distribution which is stationary at every $t \in \mathbb{R}_+$,} exists, is unique, and satisfies 
\begin{equation}
    \label{eq:AccompanyingnonHomo}
    \tilde{\rho}(t) \tilde{H}(t) = 0 \, ,
\end{equation}
making the Markov process ergodic for all times $t \in \mathbb{R}_+$.
For the trajectory $\omega$ and the initial condition $\mu_0$, we write the path probability measure of $( \tilde{X}_t )_{0 \leq t \leq T}$ as
\begin{equation}
    \label{eq:PathProbaNonHomo}
    \begin{split}
    \tilde{\mathbb{P}}_{T}(\omega) &= \\
    &\hspace{-1cm}\mu_0(x_0) \exp \left( \int_0^{\deletetext{\tau_0}{\newtext{t'_0}}} \tilde{H}_{x_0x_0}(t) \, dt \right) H_{x_0 x_1}(\deletetext{\tau_0}\newtext{t'_0}) d\deletetext{\tau_0}\newtext{t'_0} \times \\ 
    &\hspace{0.8cm}\cdots \times \exp \left( \int_{\deletetext{\tau_{n-1}}{\newtext{t'_{n-1}}}}^{T} \tilde{H}_{x_{n} x_{n}} (t) \, dt \right) \, ,
    \end{split}
\end{equation}
and assume it to be absolutely continuous with respect to the path measure of the original SIJP in \er{eq:PathProbaSIJP} for every $T \in \mathbb{R}_+$. This assumption is supported by the fact that $\mathbb{P}_T$ and $\tilde{\mathbb{P}}_T$ have the same support as no breaking of irreducibility is allowed for the SIJPs here considered as mentioned at the end of Sec.\ \ref{sec:model}. 


Having introduced the fundamental ingredients, we can now formally calculate the probability of hitting a rare event in the SIJP by an exponential change of measure. We write
\begin{equation}
    \label{eq:ProbaRareTiltingSIJP}
    \begin{split}
    P_T(\ell,\phi) &= \mathbb{E}_{\tilde{\mathbb{P}}_T} \left[ \mathbf{1}_{\ell}(L(T)) \mathbf{1}_{\phi}(\Phi(T)) \frac{d \mathbb{P}_T}{d \tilde{\mathbb{P}}_T} (\omega) \right] \\
    &\asymp \text{exp} \left[ - T I_{\text{2.5}}(\ell,\phi) \right] \, ,
    \end{split}
\end{equation}
where the first equality is simply the definition of probability multiplied and divided by the path measure \eqref{eq:PathProbaNonHomo}. By construction, the non-homogeneous Markov process typically returns the pair $(\ell,\phi)$, and the Radon--Nikodym derivative between path measures exists by absolute continuity and takes the asymptotic form in the second line. This latter expression we use to define the level-2.5 rate function $I_{\text{2.5}}(\ell,\phi)$. Such an exponential change of measure, or tilting, is widely adopted in importance sampling techniques whereby to sample an atypical event of a Markov process one introduces a new biased distribution which typically returns the event of interest~\cite{torrie1977nonphysical,Bucklew2004,chandler1987introduction,ray2018importance,klymko2018rare,Guyader2020,frenkel2023understanding} and corrects the quantity averaged with an appropriate ``umbrella''. 

\subsection{Time-scale separation, time rescaling and reversal}


We now consider the calculation of the exponential factor in \er{eq:ProbaRareTiltingSIJP}. Using the definitions in Eqs.\ \eqref{eq:PathProbaSIJP} and \eqref{eq:PathProbaNonHomo} yields
\begin{equation}
\label{eq:25Step1}
\begin{split}
        \log \frac{d\mathbb{P}_{T}}{d \tilde{\mathbb{P}}_{T}} (\omega) 
        &= \\
        &\hspace{-1.5cm} \sum_{i=0}^{n} \bigg( \int_{\deletetext{\tau_{i-1}}\newtext{t'_{i-1}}}^{\deletetext{\tau_i}\newtext{t'_{i}}} \big( Q_{x_{i} x_{i}} (A_t) - \tilde{H}_{x_{i}x_{i}}(t) \big) \, dt \bigg) + \\ 
        &\hspace{-0.5cm} \sum_{i=1}^n  \, \,\log \frac{Q_{x_{i-1} x_i}(A_{\deletetext{\tau_{i-1}}\newtext{t'_{i-1}}})}{\tilde{H}_{x_{i-1}x_i}(\deletetext{\tau_{i-1}}\newtext{t'_{i-1}})} \, ,
\end{split}
\end{equation}
where $\deletetext{\tau}\newtext{t'}_{-1} \coloneqq 0$ and $\deletetext{\tau}\newtext{t'}_n \coloneqq T$. By introducing the identities $\sum_{x \in \mathcal{S}} \mathbf{1}_x(\tilde{X}_t)$ and $\sum_{(x,y)\in \E} \mathbf{1}_x(\tilde{X}_{t_-}) \mathbf{1}_y(\tilde{X}_{t})$, and using conservation of probability in the forms of Eqs.\ \eqref{eq:DiagRateSIJP} and \eqref{eq:DiagRateNonHomo}, \er{eq:25Step1} simplifies to
\begin{equation}
    \label{eq:25Step2}
    \begin{split}
        \log \frac{d\mathbb{P}_{T}}{d \tilde{\mathbb{P}}_{T}} (\omega) 
        &= \\
        &\hspace{-1.5cm} - \sum_{(x,y) \in \E} \bigg[ \int_0^T   \mathbf{1}_x(\tilde{X}_t) \big( Q_{x y} (A_t) -\tilde{H}_{xy}(t) \big) \, dt  \\ 
        &\hspace{-0.5cm}  + \sum_{\substack{0 \leq t \leq T \\ \Delta \tilde{X}_t \neq 0}}  \,\, \mathbf{1}_x(\tilde{X}_{t_-}) \mathbf{1}_y(\tilde{X}_t)  \log \frac{\tilde{H}_{xy}(t)}{Q_{x y}(A_t)} \bigg] \, ,
\end{split}
\end{equation}
where we also used the fact that for the specific trajectory $\omega$ if the process is at $x$ at time $\deletetext{\tau}\newtext{t'}_{i-1}$, then it will stay there at all times $t \in (\deletetext{\tau}\newtext{t'}_{i-1},\deletetext{\tau}\newtext{t'}_i]$. We remark at this stage that due to the time dependence of the rate matrices it is not possible to write \er{eq:25Step2} explicitly as a function of $L(T)$ and $\Phi(T)$, which is instead possible for time-homogeneous Markov processes. However, it is still possible to simplify \er{eq:25Step2} arguing, as we do in the following, a separation of time-scales. 

On the one hand, the SIJP's inherent time-scale is set by the lowest jump rate in $Q(A_t)$. On the other hand, the rate matrix itself changes over time due to the dependence on the empirical observable in \er{eq:EmpDepSIJP} [or Eq.\  \eqref{eq:EmpDepSIJPConfig}]. At short times, the jump rates may change rapidly due to the $t^{-1}$ appearing in $A_t$. However, as time progresses, $A_t$, $L(t)$ and $\Phi(t)$ tend to stiffen: this progressive slowness is also due to the factor $t^{-1}$, which implies that to accumulate a change of a $O(1)$ a time of $O(t)$  is needed. Therefore, in the long-time limit, we identify a fast mode, inherent of the microscopic dynamics of the SIJP, and a slower mode consequence of the dependence of the rate matrix over an empirical occupation measure and flux that change slowly. 

In the following, we make this multi-scale dynamics argument more rigorous, introducing sufficient assumptions that support the derivation. We set $T \gg 1$ and split the integral over time into $K$ integrals spanning the time intervals $I_k = [t_k, t_{k+1})$ with $\Delta t = t_{k+1} - t_k$. The following sequence of inequalities holds:
\begin{equation}
\label{eq:Inequalities}
\begin{split}
    \tau_{\text{micro}} &\coloneqq \max \left\lbrace Q^{-1}_{xy}(A_t) \right\rbrace_{(x,y) \in \E} \approx \\ 
    &\approx \max \left\lbrace \tilde{H}^{-1}_{xy}(t) \right\rbrace_{(x,y) \in \E} 
    \ll\Delta t \ll t \eqqcolon \tau_{\text{slow}} \, ,
    \end{split}
\end{equation}
where $\tau_{\text{micro}}$ identifies the time-scale associated to the microscopic SIJP as well as to the auxiliary time-inhomogeneous Markov process. The time-scale separation above justifies that within a time block of size $\Delta t$, $L(t)$, $\Phi(t)$, and $A_t$ can be treated as constant, while the microscopic processes---which are themselves generated by frozen generators---mix on a characteristic time-scale $\tau_{\text{micro}}$ which is much shorter than $\Delta t$. \newtext{The sequence of inequalities above give the conditions under which we can carry out the block-averaging calculation below. The left inequality requires the frozen auxiliary chain to equilibrate within a block, while the right inequality requires both the original SIJP generator and the auxiliary generator to remain approximately constant over the same block. For the SIJP, the variation time of $Q(A_t)$ is of order $t$ as argued above. For the auxiliary generator, it is thus sufficient that its variation time is of order $t$ or larger. Auxiliary generators varying on a faster time scale are not ruled out as stochastic processes, but they are outside the frozen-block derivation used here.} 

\newtext{Block-averaging is consequence of}\deletetext{due to the} mixing over the time window $\Delta t$\deletetext{, w}\newtext{. W}e\deletetext{can} write
\begin{align}
\label{eq:FrozenAccompany1}
\int_{t_k}^{t_{k+1}}  \mathbf{1}_x(\tilde
X_t) \, dt &\approx \Delta t \, \tilde{\rho}_x(t_k) \\
\label{eq:FrozenAccompany2}
\sum_{\substack{t_k \leq t \leq t_{k+1} \\ (\tilde{X}_{t_-}, \tilde{X}_{t}) \in \E \neq 0 }} \mathbf{1}_x(\tilde{X}_{t_-}) \mathbf{1}_y(\tilde{X}_{t}) &\approx \Delta t \, \tilde{\rho}_x(t_k) \tilde{H}_{xy}(t_k) \, ,
\end{align}
while the generators $Q(A_{t_k})$ and $\tilde{H}(t_k)$ are frozen and, therefore, approximately constant. \newtext{We also stress that, as outlined in Eq.\ \eqref{eq:Inequalities}, the block-averaging argument requires the frozen SIJP generator and the frozen auxiliary generator to relax to their accompanying distributions within the time block $\Delta t$. The block size $\Delta t$ can be chosen a posteriori. Since the state space $\mathcal S$ is finite and strongly connected, and since the rates of both the SIJP and the auxiliary process are assumed to be uniformly positive on the allowed transitions, the corresponding frozen generators have spectral gaps uniformly bounded away from zero. In particular, one may assume
\begin{equation}
\label{eq:UniformaBoundAuxiliary}
\inf_{t>0} \text{gap} \, \tilde{H}(t) > 0 ,
\end{equation}
and analogously for $Q(A_t)$. This gives a uniform upper bound on the mixing times of the frozen dynamics. One can then choose $\Delta t$ larger than this bound, while still satisfying the remaining scale-separation inequalities in Eq.\ \eqref{eq:Inequalities}.}

We can therefore write
\begin{equation}
    \label{eq:25Step3}
    \begin{split}
        \log \frac{d\mathbb{P}_{T}}{d \tilde{\mathbb{P}}_{T}} (\omega) 
        &= \\
        &\hspace{-1.5cm} - \sum_{(x,y) \in \E} \bigg[ \sum_{k=0}^{K-1}   \tilde{\rho}_x(t_k) \big( Q_{x y} (A_{t_k}) -\tilde{H}_{xy}(t_k) \big) \, \Delta t  \\ 
        &\hspace{-1.2cm}  + \sum_{k=0}^{K-1}  \tilde{\rho}_x(t_k) \tilde{H}_{xy}(t_k)  \log \frac{\tilde{H}_{xy}(t_k)}{Q_{x y}(A_{t_k})} \, \Delta t \bigg] + \text{Err}\newtext{_T}(\Delta t) \, ,
\end{split}
\end{equation}
where the cumulative error $\deletetext{\text{Err}(\Delta t)}\newtext{\mathrm{Err}_T(\Delta t)}$ made in the time-scale separation approximation \deletetext{is controlled by how smoothly the local dynamics changes when the slow control variable $A_t$ changes.}\newtext{comes from two approximations: freezing the SIJP and auxiliary generators on each block, and replacing block averages by the corresponding frozen accompanying distribution. We can reformulate the time-scale separation in Eq.\ \eqref{eq:Inequalities} as requesting that the frozen cost density varies on the slow time scale, namely
\begin{widetext}
\begin{equation}
\label{eq:ErrorChange}
\left|\frac{d}{dt} \sum_{(x,y)\in\E}
\left[ \tilde{\rho}_x(t)\big(Q_{xy}(A_t) - \tilde H_{xy}(t)\big) + \tilde{\rho}_x(t) \tilde{H}_{xy}(t)
\log \frac{\tilde{H}_{xy}(t)}{Q_{xy}(A_t)} \right] \right| = O(t^{-1}) \, .
\end{equation}
\end{widetext} 
Hence, the cumulative error satisfies
\begin{equation}
\frac{|\text{Err}_T(\Delta t)|}{T}
=
O\left( \frac{\Delta t \log T}{T} + \frac{\tau_{\text{micro}}}{\Delta t} 
\right) \, .
\end{equation}
The first contribution is the error related to the Riemann sum. Since the frozen cost density changes by order $t^{-1}$ in Eq.\ \eqref{eq:ErrorChange}, the error made on a block starting at $t_k$ is proportional to the block length squared divided by $t_k$; summing these contributions over the blocks produces the logarithmic factor. The second contribution is due to the replacement of empirical occupations and fluxes by the frozen accompanying distribution. Making the replacement loses only a microscopic transient of order $\tau_{\text{micro}}$; summing over the blocks and dividing by the total time gives the second term. Therefore, choosing for instance $\Delta t=T^\alpha$ with $0<\alpha<1$, the accumulated error is subextensive, $\text{Err}_T(\Delta t)=o(T)$, and does not contribute to the large-deviation rate function. For more details on the error we refer the reader to Appendix~\ref{app:err}.}
\deletetext{In practice, we may want the maps $t \mapsto Q_{xy}(A_t)$ and $t \mapsto \tilde{H}_{xy}(t)$ to be of class $C^1$ on $\mathbb{R}_+$.}

Finally, by taking the limit $K \rightarrow \infty$, the sum can be approximated with the integral over the mesoscopic time-scale, yielding
\begin{equation}
    \label{eq:25Step4}
    \begin{split}
        \log \frac{d\mathbb{P}_{T}}{d \tilde{\mathbb{P}}_{T}} (\omega) 
        &= \\
        &\hspace{-1.5cm} - \sum_{(x,y) \in \E}\int_0^T \bigg(   \tilde{\rho}_x(t) \big( Q_{x y} (A_t) - \tilde{H}_{xy}(t) \big)  \\ 
        &\hspace{.5cm}  + \tilde{\rho}_x(t) \tilde{H}_{xy}(t) \log \frac{\tilde{H}_{xy}(t)}{Q_{x y}(A_t)} \bigg) \, dt \, .
\end{split}
\end{equation}
For compactness, we define
\begin{equation}
    \label{eq:25Functional}
    \begin{split}
         \mathcal{I}_T[\tilde{\rho},\tilde{H}] &\coloneqq\\
         &\hspace{-1cm}\frac{1}{T} \sum_{(x,y) \in \E}\int_0^T   \newtext{\tilde{\rho}_x(t)} Q_{x y} (A_t) \Gamma \left( \frac{\tilde{H}_{xy}(t)}{Q_{x y} (A_t)} \right)  \, dt \, ,
    \end{split}
\end{equation}
where $\Gamma(x) = x \, \newtext{\log} \, x - x + 1$ is the Cram\'{e}r (or relative-entropy) rate function for Poisson statistics. 

\newtext{Next, we make two changes of variable inside the time integral} of Eq.\ \eqref{eq:25Functional}\newtext{. First, we introduce the adimensional logarithmic time $s=\log (t / t_0)$, where $t_0$ is interpreted as the system’s relevant memory horizon and is then set to $1$ without loss of generality. As $t \in [0,T]$, $s \in (-\infty, \log \,T]$. Within this time scale, the empirical observable $A_t$ makes $O(1)$ changes in $O(1)$ time.}
We take the liberty to \newtext{set $t_0=1$}\deletetext{do so} because the calculation depends only on time derivatives and long-time asymptotics, while the additive constant generated by the logarithmic rescaling is irrelevant for our purposes. We stress, however, that $t_0$ may play an important role in concrete experimental settings where one wishes to test the theory developed here. In such cases, identifying the intrinsic time scales of the system under investigation would be necessary.

\deletetext{By logarithmically rescaling time and then reversing it, we obtain a version of Eq.\ \eqref{eq:25Functional} that no longer depends explicitly on $T$. More precisely, we first replace $T \to e^T$ and then perform the change of variable $t \to e^t$. In doing so, we assume time to be dimensionless. This can always be achieved by introducing an intrinsic time scale $t_0$, which can be interpreted as the most relevant past time horizon of the system, and by rescaling time accordingly. With abuse of notation, we then write $t \equiv t/t_0$ and similarly $T \equiv T/t_0$.}

\newtext{The first change of variable} \deletetext{E}\newtext{e}ventually\deletetext{, this} yields
\begin{equation}
    \label{eq:25Functional2}
    \begin{split}
         \mathcal{I}_{T}[\tilde{\rho},\tilde{H}] &\coloneqq \\
         &\hspace{-1.25cm}\deletetext{e^{-T}}\newtext{\frac{1}{T}} \sum_{(x,y) \in \E}\int_{-\infty}^{\newtext{ \log \, T}} e^\newtext{s}    \tilde{\rho}_x(e^\newtext{s})  Q_{xy} (A_{e^\newtext{s}}) \Gamma \left( \frac{\tilde{H}_{xy}(e^\newtext{s})}{Q_{x y}(A_{e^\newtext{s}})} \right) \, d\newtext{s} \, ,
    \end{split}
\end{equation}
which effectively sets the system on \deletetext{a}\newtext{the} (slower) logarithmic time-scale\deletetext{, over which $A_t$ accumulates $O(1)$ changes over in $O(1)$ times}. \deletetext{Finally, we reverse time by changing variable once again, i.e., $t \rightarrow T - t$.}

\newtext{Then, we make the change $\tau = \log \,T - s$ to measure backwards from the final logarithmic time $\log \, T$. Due to this change, the integral boundaries flip to $[0,\infty)$.} Eventually, this yields
\begin{equation}
    \label{eq:25FunctionalFinal}
    \begin{split}
         \mathcal{I}[\rho,H] &\coloneqq  \\ 
         &\hspace{-1cm}\sum_{(x,y) \in \E}\int_{0}^\infty e^{-\newtext{\tau}}   \rho_x(t) Q_{x y} (M_\newtext{\tau}) \Gamma \left( \frac{H_{xy}(\newtext{\tau})}{Q_{x y}(M_{\newtext{\tau}})} \right) \, d\newtext{\tau} \, ,
    \end{split}
\end{equation}
where we have (i) dropped $T$ in the notation $\tilde{\rho}\newtext{(T e^{-\tau})} \eqqcolon \rho\newtext{(\tau)}$, $A_{\newtext{T e^{-\tau}}} \eqqcolon M_{\newtext{\tau}}$, and $\tilde{H}_{xy}(\newtext{T e^{-\tau}}) \eqqcolon H_{xy}(\newtext{\tau})$ and (ii) also dropped the explicit dependence on $T$ from the definition of the functional $\mathcal{I}_{T} \eqqcolon \mathcal{I}$. Of course, all objects still implicitly depend on $T$, but in the long-time limit they become $T$ independent. Remarkably, in \er{eq:25FunctionalFinal} the exponential factor $e^{-\newtext{\tau}}$ acts as a \textit{discount}: as $\newtext{\tau}$ increases, the contribution of the corresponding time to the rate functional is exponentially suppressed. This aligns with the intuition that, at short physical times---corresponding to large $\newtext{\tau}$ under time reversal---the empirical measure fluctuates more rapidly and thus contributes less to the overall fluctuation cost than it does in the long-time limit. 

By multiplying and dividing the argument of $\Gamma$ by $\rho(\newtext{\tau})$, the rate functional $\mathcal{I}$ takes the equivalent form
\begin{equation}
    \label{eq:25FunctionalFinalEquivalent}
    \begin{split}
    \tilde{\mathcal{I}}[\rho,\eta] &\coloneqq \\
    &\hspace{-1cm}\sum_{(x,y) \in \E}\int_{0}^\infty e^{-\newtext{\tau}}   \rho_x(\newtext{\tau}) Q_{x y} (M_\newtext{\tau}) \Gamma \left( \frac{\eta_{xy}(\newtext{\tau})}{\rho_x(\newtext{\tau})Q_{x y}(M_{\newtext{\tau}})} \right) \, d\newtext{\tau} \, ,
    \end{split}
\end{equation}
where we denote
\begin{equation}
    \label{eq:InstantaneousDensNonHomo}
    \eta_{xy}(\newtext{\tau}) \coloneqq \rho_x(\newtext{\tau}) H_{xy}(\newtext{\tau}) \, ,
\end{equation}
the instantaneous flux of the rescaled and reverse-in-time inhomogeneous Markov process. This specific form will be important later when discussing kinetic bounds.

\subsection{Average over paths: Laplace approximation}


The rate functional obtained in \er{eq:25FunctionalFinal} (or \er{eq:25FunctionalFinalEquivalent}) is still related to a specific, albeit long, realisation of a trajectory of the non-homogeneous Markov process $( \tilde{X}_t )_{0 \leq t \leq T}$. The last step is to average over the ensemble of trajectories generated, as in \er{eq:ProbaRareTiltingSIJP}. Given the exponential scaling with time of the Radon--Nikodym derivative, the average is calculated using a Laplace approximation---or contraction principle in the large deviation jargon---which finally yields
\begin{equation}
    \label{eq:RateFunct25}
    I_{\text{2.5}}(\ell, \phi) = \inf_{(\rho, H) \in \mathcal{S}_{\ell,\phi}} \, \mathcal{I}[\rho,H] \, ,
\end{equation}
the level-2.5 rate function we are after. The infimum is over trajectories $(\rho(\newtext{\tau}),H(\newtext{\tau}))_{0 \leq \newtext{\tau} < \newtext{\infty}}$ belonging to the set $\mathcal{S}_{\ell,\phi}$, which contains all normalised densities $\rho(t)$ along with the following constraints
\begin{align}
    \label{eq:AccDistrReverse}
    \rho(\newtext{\tau}) H(\newtext{\tau}) &= 0 \\
    \label{eq:Ms}
    \begin{split}
    M_\newtext{\tau} &= e^\newtext{\tau} \int_\newtext{\tau}^\infty e^{-\newtext{\tau}'} \bigg( \sum_{x \in \mathcal{S}} f_x \rho_x(\newtext{\tau}') + \\ &\hspace{2.5cm}\sum_{(x,y) \in \E} g_{xy} \eta_{xy}(\newtext{\tau}') \bigg) \, d\newtext{\tau}'
    \end{split}\\
    \label{eq:RareDensity}
    \ell_x &= \int_0^\infty e^{-\newtext{\tau}'} \rho_x(\newtext{\tau}') \, d\newtext{\tau}' \\
    \label{eq:RareFlux}
    \phi_{xy} &= \int_0^\infty e^{-\newtext{\tau}'} \eta_{xy}(\newtext{\tau}') \, d\newtext{\tau}' \, ,
\end{align}
set to hold for all $x \in \mathcal{S}$ and all $(x,y) \in \E$. \er{eq:AccDistrReverse} is equivalent to \er{eq:AccompanyingnonHomo} rescaled and time-reversed. The same holds for $M_\newtext{\tau}$ in \er{eq:Ms} with respect to \er{eq:EmpDepSIJP} and for the flux $\phi$ in \er{eq:RareFlux} with \er{eq:EmpFluxSIJP}. 
Finally, a contraction equivalent to \eqref{eq:RateFunct25} can be written for the functional $\tilde{\mathcal{I}}[\rho,\eta]$, over the same set of constraints.



The minimum and zero, which is not necessarily unique, of the rate function \eqref{eq:RateFunct25} characterises the typical behaviour of the SIJP. As $\Gamma(1) = 0$, this is obtained by setting $H_{xy}(\newtext{\tau}) = Q_{xy}(M_\newtext{\tau})$. Such an equivalence is made possible by the fact that the mapping $Q_{xy}(M_\newtext{\tau})$ is local in time in the extended state space $\mathcal{S} \times \mathcal{P}_0(\mathcal{S}) \times \mathbb{R}_+^{d(d-1)}$. Therefore, it is always possible to find a deterministic operator $H_{xy}(\newtext{\tau}) = (Q_{xy}(M)) (\newtext{\tau}) $. In such a typical case, the constraints \eqref{eq:AccDistrReverse}--\eqref{eq:RareFlux} simplify to
\begin{align}
    \label{eq:AccDistrReverseStat}
    \bar{\rho}(\newtext{\tau}) Q(\bar{M}_\newtext{\tau}) &= 0 \\
    \label{eq:MsStat}
    \begin{split}
    \bar{M}_\newtext{\tau} &= e^\newtext{\tau} \int_\newtext{\tau}^\infty e^{-\newtext{\tau}'} \bigg( \sum_{x \in \mathcal{S}} f_x \bar{\rho}_x(\newtext{\tau}') + \\ &\hspace{2.5cm}\sum_{(x,y) \in \E} g_{xy} \bar{\eta}_{xy}(\newtext{\tau}') \bigg) \, d\newtext{\tau}'
    \end{split}\\
    \label{eq:RareDensityStat}
    \bar{\pi}_x &= \int_0^\infty e^{-\newtext{\tau}'} \bar{\rho}_x(\newtext{\tau}') d\newtext{\tau}' \\
    \label{eq:RareFluxStat}
    \bar{\varphi}_{xy} &= \int_0^\infty e^{-\newtext{\tau}'} \bar{\eta}_{xy}(\newtext{\tau}') \, d\newtext{\tau}' \, ,
\end{align}
where $\bar{\rho}(\newtext{\tau})$, $\bar{M}(\newtext{\tau})$, and $\bar{\eta}(\newtext{\tau})$ are the (rescaled and reverse-in-time) typical trajectories of instantaneous density, empirical observable and instantaneous flux of the original SIJP. Moreover, $(\bar{\pi},\bar{\varphi})$ are interpreted as weighted averages over the set of stationary limiting pairs $\mathcal{A}_{\mathrm{stat}}$. In the case this set is singleton, viz.\ $\mathcal{A}_{\text{stat}} = \left\lbrace (\pi,\varphi) \right\rbrace$, then $\bar{\pi} \equiv \pi$ and $\bar{\varphi} \equiv \varphi$. We denote the set of trajectories $(\bar{\rho}(\newtext{\tau}),Q(\bar{M}_\newtext{\tau}))$ satisfying Eqs.\ \eqref{eq:AccDistrReverseStat}--\eqref{eq:RareFluxStat} as $\mathcal{S}_{\mathcal{A}_{\text{stat}}}$.

We conclude this section with a few remarks. First, we can read Eqs.\ \eqref{eq:RateFunct25} and \eqref{eq:25FunctionalFinal}, along with the constraints \eqref{eq:AccDistrReverse}--\eqref{eq:RareFlux}, as a dynamical re-weighting of the relative entropy of a Poisson process---the latter given by $\Gamma(x)$---and therefore as a generalisation of the level-2.5 LDs for Markov jump processes~\cite{de2001large,maes2008canonical,bertini2012large,barato2015a-formal,chetrite2015variational}, which follows the seminal work of Donsker and Varadhan~\cite{donsker1975ii-asymptotic}. Differently from the latter however, the level-2.5 LDs for SIJPs is not, generally speaking, convex. This is a consequence of the contraction appearing in \er{eq:RateFunct25}, which may give rise to interesting dynamical behaviour in generating fluctuations in SIJPs. \deletetext{Although this should be true from a general point of view, all the examples treated in Sec.~\ref{sec:Ex} do not show non-convexity properties and more work is needed to find examples which show it.}\newtext{One example in Sec.~\ref{sec:Ex} shows interesting features. However, more work is needed to better understand the dynamical implications of having a non-convex rate function for SIJPs.}

\newtext{Although we believe we have stated all necessary assumptions to derive the large deviation principle in Eq.\ \eqref{eq:ProbaRareTiltingSIJP} along with the rate function \eqref{eq:RateFunct25}, }\deletetext{The validity of the result in \er{eq:RateFunct25} relies on a}\newtext{one should check the validity of these assumptions}\deletetext{set of assumptions that may not always hold and should therefore be verified} on a case-by-case basis---an issue commonly encountered when working formally with tilting methods. For the broad class of affine rate matrices $Q(L(t))$, however, our findings are supported by a rigorous derivation recently presented in Ref.~\cite{budhiraja2025jump}, where variational and control-theoretic formulations of Laplace functionals of Poisson random measures are used to establish level-2.5 LDs for self-interacting jump processes, see Eqs.~(6.1)–(6.2) in Ref.~\cite{budhiraja2025jump}. While the approach adopted here is less rigorous, it extends beyond affine maps. Generalising the rigorous derivations of~\cite{budhiraja2025jump} to the broader class of rate maps considered in this work and in our Ref.~\cite{coghi2026level} also remains an important direction for future research.

Finally, we stress that the general level-2.5 LDs we derived is useful in practice for two reasons. The first is that long-time LDs of any time-extensive trajectory observable, such as activity or a current, as well as the empirical measure itself, can be obtained by contraction from the LDs of empirical occupation measure and flux. The second is that \er{eq:RateFunct25} is variational, so that even if the exact minimisation is difficult, it is still possible to get suboptimal solutions as upper bounds to the true rate function, as we will show in the next two sections.

\section{Kinetic bounds}
\label{sec:Kin}


We focus on the generalised sample-mean flux observable
\begin{equation}
    \label{eq:GenFluxSIJP}
    B_T = \sum_{\substack{(x,y) \in \E}} \beta_{xy} \Phi_{xy}(T) \, ,
\end{equation}
\deletetext{where}\newtext{and} $\beta_{xy} \in \mathbb{R}_+$ and $\Phi_{xy}(T)$ is defined in \er{eq:EmpFluxSIJP}. In the following, we assume a unique stationary pair such that $\mathcal{A}_{\text{stat}} = \left\lbrace (\pi,\varphi) \right\rbrace$. Typically, $B_T \rightarrow \bar{b}$ in probability, where
\begin{equation}
    \label{eq:StatbSIJP}
    \bar{b} \coloneqq  \sum_{\substack{(x,y) \in \E}} \beta_{xy} \varphi_{xy} \, ,
\end{equation}
where $\varphi \in \mathbb{R}_+^{d \times (d-1)}$ is the typical behaviour of the empirical flux and is defined in \er{eq:RareFluxStat}. 

Here, we are interested in the fluctuations of $B_T$ in \er{eq:GenFluxSIJP}.  Its large deviation rate function $I(b)$ can be obtained by solving the variational problem in \er{eq:RateFunct25} along with the usual constraints \eqref{eq:AccDistrReverse}--\eqref{eq:RareFlux} and the extra constraint
\begin{equation}
    \label{eq:FluctbSIJP}
    b = \sum_{\substack{(x,y) \in \E}} \beta_{xy} \phi_{xy} \, ,
\end{equation}
which we express as follows for ease of readability
\begin{equation}
    \label{eq:RateIb2}
    I(b) = \inf_{\substack{(\rho,H) \in \mathcal{S}_{\ell,\phi} \\ \ell, b: \, \text{Eq.\ } \eqref{eq:FluctbSIJP}}} \mathcal{I}[\rho, H] \, .
\end{equation}

The minimum and zero of such a rate function is found at $b = \bar{b}$, by setting $\ell = \pi$, $\phi =\varphi$, and $(\rho(\newtext{\tau}),H(\newtext{\tau})) = (\bar{\rho}(\newtext{\tau}), \bar{H}(\newtext{\tau})) \in \mathcal{S}_{\pi,\varphi}$ as solutions of Eqs.\ \eqref{eq:AccDistrReverseStat}--\eqref{eq:RareFluxStat}.


\subsection{SIJP Kinetic Uncertainty Relation}
\label{subsec:KUR}


In this Subsection, and similarly when deriving the SIJP-TUR below, we restrict to the case $g=0$, namely the empirical observable dependence in the SIJP is only given by state contributions, i.e.,
\begin{equation}
    \label{eq:RestrictionKURTUR}
    A_t = t^{-1} \int_0^t f_{X_{t'}} \, dt' \, .
\end{equation}
Such a restriction allows us to derive explicit formulas and link them to well-known results for time-homogeneous Markov processes.

Although an exact minimisation of \er{eq:RateIb2} is out of reach, we can find a suboptimal solution $I^*(b)$ with a simple Ansatz where the density coincides with the stationary state, but the flux is a scalar rescaling of the typical one, that is, 
\begin{align}
    \label{eq:KURSubOptDens}
    \ell^*_x &= \pi_{x} \\
    \label{eq:KURSubOptFlux}
    \phi^*_{xy} &= \left( b/\bar{b} \right) \, \varphi_{xy} \, .
\end{align} 
This in turn implies 
\begin{equation}
\label{eq:AssKURSIJP}
(\rho^*(\newtext{\tau}),H^*(\newtext{\tau})) = (\bar{\rho}(\newtext{\tau}), (b/\bar{b}) Q(\bar{M}_\newtext{\tau})) \in \mathcal{S}_{\pi,(b/\bar{b}) \varphi} \, ,    
\end{equation} 
such that the constraints \eqref{eq:AccDistrReverseStat}--\eqref{eq:RareFluxStat} are satisfied. Notice that if we had allowed for jump type contributions in the SIJP dependence on the empirical observable, we would have not been able to write the equation above due to $M^*_\newtext{\tau} (\neq \bar{M}_\newtext{\tau})$ including a dependence on the fluctuation $b$ via $\eta^*_{xy}(\newtext{\tau}) = (b/\bar{b}) \bar{\eta}_{xy}(\newtext{\tau})$.

The meaning of the Ansatz above is clear: the time-dependent generator of the time-inhomogeneous auxiliary process is simply the generator of the SIJP rescaled by an overall factor determined by the fluctuation of interest. Eventually, \er{eq:RateIb2} reduces to
\begin{equation}
    \label{eq:IbBoundSIJP}
    I(b) \leq I^*(b) \coloneqq k_{\text{SIJP}} \Gamma \left( b / \bar{b} \right)  \, , 
\end{equation}
where
\begin{equation}
    \label{eq:DynActSIJP}
    \begin{split}
    k_{\text{SIJP}} &\coloneqq \sum_{\substack{(x,y) \in \E}} \varphi_{xy} \\
    &= \sum_{\substack{(x,y) \in \E}} \int_0^\infty e^{-\newtext{\tau}} \bar{\eta}_{xy}(\newtext{\tau}) \, d\newtext{\tau} \, ,
    \end{split}
\end{equation}
is the hereby-defined average \textit{SIJP dynamical activity}. It corresponds to the number of configuration changes per unit time and it generalises the standard dynamical activity of a time-homogeneous Markov jump process to the non-Markov case of SIJPs by adding an exponential discount in time.


We now expand $I(b)$ and $\Gamma(b/\bar{b})$ up to second order on, respectively, the l.h.s.\ and r.h.s.\ of \er{eq:IbBoundSIJP}. Then, by recalling the definition of asymptotic variance 
\begin{equation}
    \label{eq:AsymptVar}
    (I''(b))^{-1} = \text{var}(b) \, ,
\end{equation}
we obtain
 \begin{equation}
    \label{eq:KURSIJP}
    \frac{\text{var}(b)}{\bar{b}^2} \geq \frac{1}{k_\text{SIJP}} \, ,
\end{equation}
which is what we refer to as SIJP-KUR, generalising the well-known KUR for time-homogeneous Markov processes~\cite{garrahan2017simple,terlizzi2019kinetic}. Evidently, as for Markov processes, the precision of flux (or current) observables is bounded by the average dynamical activity of SIJPs.

\newtext{We remark that the identification of the inverse curvature with the asymptotic variance in Eq.\ \eqref{eq:AsymptVar} is understood in the fluctuation regime in which the observable satisfies a central limit theorem,
\begin{equation}
\sqrt{T}(B_T-\bar b)\rightarrow \mathcal N(0,\text{var}(b)) \, ,    
\end{equation}
where convergence is understood in distribution. For SIJPs, as well as for any other processes, this is not automatic. We believe that the assumptions considered so far in Sec.\ \ref{sec:model} should be enough to prove a central limit theorem in the case of a SIJP with a unique hyperbolically stable attractor. However, this goes beyond the scope of this work and leave the endeavor to future research.}

In principle, it is possible to improve the bound in \er{eq:IbBoundSIJP} by exploiting the inner time dependency of the rate functional $\mathcal{I}[\rho,H]$ and, therefore, by considering a time-dependent rescaling of the suboptimal rate matrix in \er{eq:AssKURSIJP} as 
\begin{equation}
\label{eq:AssKURTimeDepSIJP}
(\rho^*(\newtext{\tau}),H^*(\newtext{\tau})) = (\bar{\rho}(\newtext{\tau}),\alpha_\newtext{\tau} Q(\bar{M}_\newtext{\tau})) \in \mathcal{S}_{\pi, \phi^*} \, ,    
\end{equation}
where $\phi^*_{xy} = \int_0^\infty e^{-\newtext{\tau}} \alpha_\newtext{\tau} \bar{\eta}_{xy}(\newtext{\tau})$ cannot be cast into an explicit function of the typical flux $\varphi_{xy}$ as in Eq.\ \eqref{eq:KURSubOptFlux}. The new bound takes the form 
\begin{equation}
    \label{eq:IbBoundAlphaTimeSIJP}
    I(b) \leq I^{**}(b) \coloneqq \sum_{\substack{(x,y) \in \E}} \int_0^\infty e^{-\newtext{\tau}} \Gamma(\alpha_\newtext{\tau}) \bar{\eta}_{xy}(\newtext{\tau}) \, d\newtext{\tau} \, .
\end{equation}
By selecting specific trajectories of  $\alpha_\newtext{\tau}$ it should be possible to obtain $I^{**}(b) \leq I^*(b)$. In particular, one could look after the best solution for $\alpha_\newtext{\tau}$ such that $I^{**}(b)$ is minimal. Variationally, one can show that such an optimal $\alpha^*_\newtext{\tau}$ satisfies the closed form 
\begin{equation}
    \label{eq:OptAlphaSIJP}
    \alpha_\newtext{\tau}^* = e^{ - \lambda \frac{\sum_{(x,y)\in \E} \beta_{xy} \bar{\eta}_{xy}(\newtext{\tau})}{\sum_{(x,y) \in \E} \bar{\eta}_{xy}(\newtext{\tau})}} \, ,
\end{equation}
where the Lagrange multiplier $\lambda$ plays the role of a \textit{temperature} or \textit{external parameter} and is fixed by the (now dynamical) constraint
\begin{equation}
    \label{eq:Lagrange}
    b = \sum_{\substack{(x,y) \in \E}} \beta_{xy} \int_0^\infty e^{-\newtext{\tau}} \alpha_\newtext{\tau}^*   \bar{\eta}_{xy}(\newtext{\tau}) \, d\newtext{\tau} \, .
\end{equation}
The stationary constraint $\bar{\rho}_\newtext{\tau} \alpha_\newtext{\tau} Q(\bar{M}_\newtext{\tau}) = 0$ selects the rescaling $\alpha_\newtext{\tau}$ such that it admits $\bar{\rho}_\newtext{\tau}$ as the instantaneous stationary distribution of $\alpha_\newtext{\tau} Q(\bar{M}_\newtext{\tau})$ as required by \eqref{eq:AccDistrReverse} for $(\rho^*(\newtext{\tau}),H^*(\newtext{\tau}))$ and is automatically satisfied as long as $\alpha_\newtext{\tau} \neq 0$ for every $\newtext{\tau} \in \mathbb{R}_+$. However, as mentioned earlier, for the time-scale separation and the level-2.5 LDs to hold we need $\tilde{H}(\newtext{\tau})$ to (i) change slowly (cumulate $O(1)$ changes over in $O(t)$ time or longer) and (ii) smoothly. As a consequence, this further restricts the class of functions $\alpha_\newtext{\tau}$ we can admit for the bound in \er{eq:IbBoundAlphaTimeSIJP} to hold.


\subsection{SIJP Ultimate Kinetic Uncertainty Relation}
\label{subsec:UKUR}

In this Subsection, in order to obtain an explicit result, we restrict ourselves to the---opposite with respect to the previous Subsection---case $f = 0$, namely, the empirical observable dependence in the SIJP is given by
\begin{equation}
    A_t = t^{-1} \sum_{\substack{0 \leq t' \leq t \\ (X_{t'_-}, X_{t'}) \in \E}} g_{X_{t'_-},X_{t'}} \, ,
\end{equation}
which includes only jump type contributions.

By rephrasing the variational problem in \er{eq:RateIb2} in terms of the functional $\tilde{\mathcal{I}}[\rho,\eta]$ in \er{eq:25FunctionalFinalEquivalent} we obtain 
\begin{equation}
    \label{eq:RateIb2Equivalent}
    I(b) = \inf_{\substack{(\rho,\eta) \in \mathcal{S}_{\ell,\phi} \\ \ell, b: \, \text{Eq.\ } \eqref{eq:FluctbSIJP}}} \tilde{\mathcal{I}}(\rho, \eta) \, .
\end{equation}

A suboptimal solution to the display above can be found by admitting the Ansatz
\begin{align}
    \label{eq:AssRhosUKURSIJP}
    \rho^*_x(\newtext{\tau}) &= \bar{\rho}_x(\newtext{\tau}) + \epsilon \, \delta_x \\
    \label{eq:AssEtasUKURSIJP}
    \eta^*_{xy}(\newtext{\tau}) &= (1 + \epsilon \, \Delta) \bar{\eta}_{xy}(\newtext{\tau}) \, ,
\end{align}
which generalises the approach of Ref.~\cite{macieszczak2024ultimate} to our non-Markovian setting. Hence, from \er{eq:Ms} we also have
\begin{equation}
    \label{eq:AssMsUKURSIJP}
    M^*_\newtext{\tau} = \bar{M}_\newtext{\tau} + \epsilon \, \Delta \, \delta m_\newtext{\tau} \, ,
\end{equation}
with 
\begin{equation}
    \label{eq:Assdeltam}
    \delta m_\newtext{\tau} =  e^\newtext{\tau} \int_\newtext{\tau}^\infty e^{-\newtext{\tau}'} \sum_{(x,y) \in \E} g_{xy} \bar{\eta}_{xy}(\newtext{\tau})  \, ,
\end{equation}
where $\epsilon \ll 1$, $\sum_{x \in \mathcal{S}} \delta_x = 0$ as $\rho^*(\newtext{\tau})$ is normalised, and $\Delta \in \mathbb{R}_+$. As a consequence, \er{eq:FluctbSIJP} takes the form
\begin{equation}
    \label{eq:AssbUKURSIJP}
    b = (1+\epsilon \, \Delta) \bar{b} \, ,
\end{equation}
with $\bar{b}$ given in \er{eq:StatbSIJP}. We notice that the assumption on the rate matrix of the non-homogeneous Markov process in \er{eq:AssKURSIJP}, which led to the SIJP-KUR, is fundamentally different from the assumption in \er{eq:AssEtasUKURSIJP}. It is such because the former allows for fluctuations in the density to influence fluctuations in the flux, which is not true for the latter. [Notice that the SIJP-KUR can also be derived following the steps below by assuming \er{eq:AssRhosUKURSIJP} and $H^*_{xy}(\newtext{\tau}) = (1 + \epsilon \, \Delta) Q_{xy}(\bar{M}_\newtext{\tau})$ in place of \er{eq:AssEtasUKURSIJP}, with $b/\bar{b} = 1 + \epsilon \Delta$.]


Using Eqs.\  \eqref{eq:AssRhosUKURSIJP}--\eqref{eq:AssMsUKURSIJP} and expanding up to second order in $\epsilon$, \er{eq:25FunctionalFinalEquivalent} yields
\begin{equation}
    \label{eq:Funct2WithAssSIJP}
    \begin{split}
    \tilde{\mathcal{I}}[\rho^*,\eta^*] &= \frac{\epsilon^2}{2} \sum_{y \in \mathcal{S}} \int_0^\infty \left( e^{-\newtext{\tau}} \frac{Q_{xy}(\bar{M}_\newtext{\tau})}{\bar{\rho}_x(\newtext{\tau})} \tilde{\delta}_{xy}^2(\newtext{\tau}) \right) \, dt \\ 
    &\hspace{2cm} + O(\epsilon^3) \, ,
    \end{split}
\end{equation}
where we define 
\begin{equation}
    \tilde{\delta}_{xy}(\newtext{\tau}) \coloneqq \delta_x - \Delta \, \bar{\rho}_x(\newtext{\tau}) \left( 1 - \frac{\delta m_\newtext{\tau} \, Q'_{xy}(\bar{M}_\newtext{\tau})}{Q_{xy}(\bar{M}_\newtext{\tau})} \right) \, ,
\end{equation} 
with $Q'_{xy}(\bar{M}_\newtext{\tau}) \coloneqq d/dM \, Q_{xy}(M)|_{M = \bar{M}_\newtext{\tau}}$, and such that 
\begin{equation}
\sum_{x \in \mathcal{S}} \tilde{\delta}_{xy}(\newtext{\tau}) = - \Delta \left( 1 + \delta m_\newtext{\tau} \sum_{x \in \mathcal{S}} \frac{\bar{\rho}_x(\newtext{\tau}) Q'_{xy}(\bar{M}_\newtext{\tau})}{Q_{xy}(\bar{M}_\newtext{\tau})} \right) \, ,
\end{equation}
for all $\newtext{\tau} \in \mathbb{R}_+$. By a straightforward application of the Cauchy--Schwarz inequality in the form 
\begin{equation}
    \label{eq:CauchySchwarz}
    \left( \sum_{x \in \mathcal{S}} a_{xy}(\newtext{\tau})^{-1} \right) \left( \sum_{x \in \mathcal{S}} a_{xy}(\newtext{\tau}) b_{xy}^2(\newtext{\tau}) \right) \geq \left( \sum_{x \in \mathcal{S}} b_{xy}(\newtext{\tau}) \right)^2 \, ,
\end{equation}
where we identify $a_{xy}(\newtext{\tau}) = Q_{xy}(\bar{M}_\newtext{\tau})/\bar{\rho}_x(\newtext{\tau})$ and $b_{xy}(\newtext{\tau}) = \tilde{\delta}_{xy}(\newtext{\tau})$, and remark that the inequality is saturated for 
\begin{equation}
    \label{eq:CauchySchwarzSaturation}
    b_{xy}(\newtext{\tau}) = \frac{\sum_{x \in \mathcal{S}} b_{xy}(\newtext{\tau})}{a_{xy}(\newtext{\tau}) \sum_{x \in \mathcal{S}} a^{-1}_{xy}(\newtext{\tau})} \, , 
\end{equation}
\er{eq:Funct2WithAssSIJP} simplifies to 
\begin{equation}
        \label{eq:BoundRateFuncUKUR}
        \begin{split}
        \tilde{\mathcal{I}}(\rho^*,\eta^*) &= \frac{\epsilon^2 \Delta^2}{2} \sum_{y \in \mathcal{S}} \Bigg( \\
        &\hspace{-1.5cm} \int_0^\infty e^{-\newtext{\tau}} \frac{\left( 1 - \delta m_\newtext{\tau} \sum_{x \in \mathcal{S}} \frac{\bar{\rho}_x(\newtext{\tau}) Q'_{xy}(\bar{M}_\newtext{\tau})}{Q_{xy}(\bar{M}_\newtext{\tau})} \right)^2}{\sum_{x \in \mathcal{S}} \frac{\bar{\rho}_x(\newtext{\tau})}{Q_{xy}(\bar{M}_\newtext{\tau})}} d\newtext{\tau} \Bigg) \\
        &\hspace{-1cm}+ O(\epsilon^3) \, .
    \end{split}
\end{equation}

Inverting \er{eq:AssbUKURSIJP} to extract $\epsilon \, \Delta$ as a function of $b$ and $\bar{b}$ and replacing this into \er{eq:BoundRateFuncUKUR} finally yields the quadratic rate function bound
\begin{equation}
    \label{eq:FinalUKURRateBound}
    \begin{split}
    I(b) &\leq I^\triangle(b) \coloneqq \frac{(b- \bar{b})^2}{2 \bar{b}^2} \times         \\ &\hspace{-0.5cm} \sum_{y \in \mathcal{S}} \Bigg( \int_0^\infty e^{-\newtext{\tau}} \frac{\left( 1 - \delta m_\newtext{\tau} \sum_{x \in \mathcal{S}} \frac{\bar{\rho}_x(t) Q'_{xy}(\bar{M}_\newtext{\tau})}{Q_{xy}(\bar{M}_\newtext{\tau})} \right)^2}{\sum_{x \in \mathcal{S}} \frac{\bar{\rho}_x(\newtext{\tau})}{Q_{xy}(\bar{M}_\newtext{\tau})}} d\newtext{\tau} \Bigg) \, .
    \end{split}
\end{equation}

Eventually, by expanding $I(b)$ in \er{eq:RateIb2Equivalent} up to second order in $\epsilon$, using the definition of asymptotic variance in \er{eq:AsymptVar}, and comparing the form obtained to \er{eq:BoundRateFuncUKUR} we obtain
\begin{equation}
    \label{eq:UKURSIJP}
    \begin{split}
    \frac{\text{var}(b)}{\bar{b}^2} &\geq \\
    &\hspace{-1.cm} \left( \sum_{y \in\mathcal{S}}\int_0^\infty e^{-\newtext{\tau}} \frac{\left( 1 - \delta m_\newtext{\tau} \sum_{x \in \mathcal{S}} \frac{\bar{\rho}_x(\newtext{\tau}) Q'_{xy}(\bar{M}_\newtext{\tau})}{Q_{xy}(\bar{M}_\newtext{\tau})} \right)^2}{\sum_{x \in S} \frac{\bar{\rho}_x(\newtext{\tau})}{Q_{xy}(\bar{M}_\newtext{\tau})}} d\newtext{\tau} \right)^{-1} \, .
    \end{split}
\end{equation}
This formula states that the asymptotic precision of the generalised flux $B_T$ is controlled by a dynamically discounted average lifetime in the instantaneous---valid for all $\newtext{\tau} \in \mathbb{R}_+$---steady state. \er{eq:UKURSIJP} extends the UKUR valid for Markov jump processes~\cite{macieszczak2024ultimate} (see also \cite{prech2024optimal}) to non-Markov SIJPs. It is immediate to check that for $A_t = 0$---this eliminates the functional dependence of the rate matrix upon the empirical observable---$\delta m_\newtext{\tau} = 0$ and all temporal dependences disappear, retrieving the usual UKUR \cite{macieszczak2024ultimate,prech2024optimal}. 

We conclude by remarking that although in the Markov case UKUR is tighter than KUR---the weighted harmonic mean is greater than the corresponding weighted arithmetic mean---in
 the SIJP process case it is no longer a general consequence.


\section{Thermodynamic bounds}
\label{sec:Thermo}


In this Section, we turn the focus on a subset of observables of the generalised flux $B_T$ in \er{eq:GenFluxSIJP}, namely generalised current observables of the form
\begin{equation}
\label{eq:GenCurrSIJP}
J_T = \sum_{x < y} \beta_{xy} \mathfrak{J}_{xy}(T) \, ,
\end{equation}
where we denote the edge currents
\begin{equation}
\label{eq:CurrsSIJP}
\mathfrak{J}_{xy}(T) = \Phi_{xy}(T) - \Phi_{yx}(T) \, .
\end{equation} 
The generalised current observable in \eqref{eq:GenCurrSIJP} is obtained from $B_T$ in \er{eq:GenFluxSIJP} by setting $\beta_{xy} = - \beta_{yx}$ for pairs $(x,y) \in \E$. Restricting to these observables allows us to refine the SIJP-KUR in \er{eq:KURSIJP}.

In order to derive explicit results as done in the SIJP-KUR case, we assume a unique stationary pair $\mathcal{A}_{\text{stat}} = \left\lbrace (\pi,\varphi) \right\rbrace$ and restrict the type of SIJP dependence we consider to \er{eq:RestrictionKURTUR}, that is, only empirical state contributions are allowed to influence the SIJP dynamics. \newtext{Furthermore, similarly to the kinetic bounds presented in the previous Section, we stress here as well that the thermodynamic bound presented below should be understood as restricted to the limit of validity of the central limit theorem for SIJPs.}

First, we derive a large deviation rate function for the edge current observable in \er{eq:CurrsSIJP}, opportunely contracting the level-2.5 large deviation rate function \eqref{eq:RateFunct25}---with $\tilde{\mathcal{I}}[\rho,\eta]$ in \er{eq:25FunctionalFinalEquivalent}. For notation purposes, it will be useful to refer to edge current fluctuations using the rescaled and time-reversed characterisation of the flux $\phi$ in \er{eq:RareFlux} and write
\begin{align}
            \label{eq:CurrRecastTURSIJP0}
            \mathfrak{j}_{xy} &= \phi_{xy} - \phi_{yx} \\ 
\label{eq:CurrRecastTURSIJP}            
            &= \int_0^\infty e^{-\newtext{\tau}} v_{xy}(\newtext{\tau}) \, d\newtext{\tau}
             \, ,
\end{align}
where we denote 
\begin{equation}
\label{eq:InstCurrTUR}
v_{xy}(\newtext{\tau}) = \eta_{xy}(\newtext{\tau}) - \eta_{yx}(\newtext{\tau}) \, ,
\end{equation}
as the antisymmetric part of the time-dependent flux. 

\subsection{SIJP Entropy Bound on Edge Currents}


We perform a first contraction of the rate functional $\tilde{\mathcal{I}}[\rho,\eta]$ over all possible instantaneous fluxes $\eta(\newtext{\tau})$ such that the constraint \eqref{eq:InstCurrTUR} is satisfied along with the conservation of probability in \er{eq:AccDistrReverse}. To do so, we introduce the extended Lagrangian
\begin{equation}
                \label{eq:LagrangianTUR}
                \begin{split}
                \mathcal{L}[\eta, \zeta, \varepsilon] &\coloneqq \\
 &\hspace{-1.5cm}\sum_{(x,y) \in \E}\int_{0}^\infty e^{-\newtext{\tau}}   \rho_x(\newtext{\tau}) Q_{x y} (M_\newtext{\tau}) \Gamma \left( \frac{\eta_{xy}(\newtext{\tau})}{\rho_x(\newtext{\tau})Q_{x y}(M_{\newtext{\tau}})} \right) \, d\newtext{\tau} \\
&\hspace{-1.2cm} + \int_0^\infty e^{-\newtext{\tau}} \zeta_{xy}(\newtext{\tau}) \left( v_{xy}(\newtext{\tau}) - \eta_{xy}(\newtext{\tau}) + \eta_{yx}(\newtext{\tau}) \right) \, d\newtext{\tau} \\ 
&\hspace{-1.2cm} + \int_0^\infty e^{-\newtext{\tau}} \varepsilon_y(\newtext{\tau}) \left( \sum_{x \in \mathcal{S}} \eta_{xy}(\newtext{\tau}) \right) \, d\newtext{\tau} \, ,
                \end{split}
\end{equation}
which includes the fields $\zeta(\newtext{\tau})$ and $\varepsilon(\newtext{\tau})$ fixing the respective constraints. We notice that the symmetry inherent in the definition of flux $\sum_x \eta_{xy}(\newtext{\tau}) = - \sum_x \eta_{yx}(\newtext{\tau})$ is inherited by the corresponding Lagrangian field, which then satisfies the relation $\varepsilon_y(\newtext{\tau}) = - \varepsilon_x(\newtext{\tau})$. Additionally, we denote $\triangle_{xy}(\newtext{\tau}) = \exp(\zeta_{xy}(\newtext{\tau}) - \zeta_{yx}(\newtext{\tau}))$. Taking variations of $\eta_{xy}(\newtext{\tau})$ over the extended Lagrangian in \er{eq:LagrangianTUR}, the optimal flux can be explicitly written in terms of its asymmetric component, the time-dependent current $v(\newtext{\tau})$, and its symmetric component, the time-dependent traffic $u(\newtext{\tau})$, as
\begin{equation}
                \label{eq:FluxCurrTURSIJP}
                \eta^*_{xy}(\newtext{\tau}) 
                    = \frac{v_{xy}(\newtext{\tau}) + u_{xy}(\newtext{\tau})}{2} \, ,                             
\end{equation}
where the traffic is explicitly given by
\begin{equation}
\label{eq:TraffTURSIJP}
\begin{split}
u_{xy}(\newtext{\tau}) =  \sqrt{v_{xy}^2(\newtext{\tau}) + 4 \rho_x(\newtext{\tau}) \rho_y(\newtext{\tau}) Q_{xy}(M_\newtext{\tau}) Q_{yx}(M_\newtext{\tau})} \, .
\end{split}
\end{equation}

Eventually, replacing $\eta^*(\newtext{\tau})$ from \er{eq:FluxCurrTURSIJP} into the rate functional $\tilde{\mathcal{I}}[\rho,\eta]$ and using the relation between ``$\log$" and ``$\arcsinh$" functions, we can write
\begin{equation}
            \label{eq:RateCurrTURSIJPTraffic}
            \begin{split}
            \tilde{\tilde{\mathcal{I}}}[\rho,v] &\coloneqq \sum_{x < y} \int_0^\infty e^{-\newtext{\tau}} \Psi(v_{xy}(\newtext{\tau}),v_{xy}^\rho(\newtext{\tau}),u_{xy}^\rho(\newtext{\tau})) \, d\newtext{\tau} \, ,
            \end{split}
        \end{equation}
with the definitions
\begin{align*}
            v_{xy}^\rho(\newtext{\tau}) &\coloneqq \rho_x(\newtext{\tau})  Q_{xy}(M_\newtext{\tau}) - \rho_y(\newtext{\tau}) Q_{yx}(M_\newtext{\tau}) \\
            u_{xy}^\rho(\newtext{\tau}) &\coloneqq \rho_x(\newtext{\tau})  Q_{xy}(M_\newtext{\tau}) + \rho_y(\newtext{\tau}) Q_{yx}(M_\newtext{\tau}) \\
            \Psi(p,q,r) &\coloneqq p \arcsinh \frac{p}{\sqrt{r^2-q^2}} - p \arcsinh \frac{q}{\sqrt{r^2-q^2}} \\
            &\hspace{-0cm}- \left(\sqrt{p^2-q^2+r^2} -r \right) \, ,
        \end{align*}
where the last form also appears in the time-homogeneous Markov case~\cite{gingrich2016dissipation} (without time dependence, of course). 


Thanks to the explicit characterisation of the edge traffic that comes naturally from the minimisation above, we can now write the rate function of the edge currents as
\begin{equation}
            \label{eq:CurrRateSimplTURSIJP1}
            \begin{split}
            I(\mathfrak{j}) &=
            \\
            &\hspace{-0.5cm}
            \inf_{\substack{ (\ell, \rho(\newtext{\tau}),v(\newtext{\tau})) \\ \ell: \text{Eqs.\ } \eqref{eq:Ms},\eqref{eq:RareDensity} \\ \mathfrak{j}: \text{Eq.\ } \eqref{eq:CurrRecastTURSIJP}}} \sum_{x < y} \int_0^\infty e^{-\newtext{\tau}} \Psi(v_{xy}(\newtext{\tau}),v_{xy}^\rho(\newtext{\tau}),u_{xy}^\rho(\newtext{\tau})) \, d\newtext{\tau} \, .
            \end{split}
\end{equation}

This object is easily bounded from above using a quadratic function obtained from considering as a suboptimal solution for the asymptotic density the stationary density, i.e.,
\begin{align}
            \label{eq:piTURSIJP}
            \ell^*_x &= \pi_x \, ,
        \end{align}
which, in turn, implies $\rho^*(\newtext{\tau}) = \bar{\rho}(\newtext{\tau})$. Then, \er{eq:CurrRateSimplTURSIJP1} simplifies to
\begin{equation}
            \label{eq:CurrRateUppTURSIJP}
            \begin{split}
            I(\mathfrak{j}) &\leq \\
            &\hspace{-0.85cm}\inf_{\substack{v(\newtext{\tau} )\\ \mathfrak{j}: \text{ Eq.\ } \eqref{eq:CurrRecastTURSIJP} }} \sum_{x < y} \int_0^\infty e^{-\newtext{\tau}} \frac{(v_{xy}(\newtext{\tau}) - v_{xy}^{\bar{\rho}}(\newtext{\tau}))^2}{4 (v_{xy}^{\bar{\rho}}(\newtext{\tau}))^2} \sigma_{xy}^{\bar{\rho}}(\newtext{\tau}) \, d\newtext{\tau} \, ,
            \end{split}
        \end{equation}
where we interpret
\begin{equation}
\label{eq:InstEdgeEntrProdSIJP}
\sigma_{xy}^{\bar{\rho}}(\newtext{\tau}) \coloneqq 2 v_{xy}^{\bar{\rho}}(\newtext{\tau}) \arcsinh \frac{v_{xy}^{\bar{\rho}}(\newtext{\tau})}{\sqrt{\big(u_{xy}^{\bar{\rho}}(\newtext{\tau})\big)^2-\big(v_{xy}^{\bar{\rho}}(\newtext{\tau})\big)^2}} \, ,
\end{equation}
as the SIJP instantaneous edge entropy production. \newtext{\newtext{This interpretation is valid because the empirical occupation measure evolves slowly: the generator $Q(\bar{M}_\tau)$ changes quasi-statically, so that the underlying SIJP can equilibrate at each frozen time and $\bar{\rho}(\tau)$ is stationary with respect to $Q(\bar{M}_\tau)$~\cite{Jarzynski1997,Seifert2018}.}} \deletetext{This interpretation holds as $\bar{\rho}(\newtext{\tau})$ is stationary for $Q(\bar{M}_\newtext{\tau})$ and, in particular, $\bar
M_\newtext{\tau}$ can be thought of as an ``internal" adiabatic protocol.}Finally, to get rid of the last minimisation, one could set the suboptimal $v^*(\newtext{\tau}) = \mathfrak{j}$ (constant over time) or replace the optimal minimising instantaneous current, which takes the explicit form
\begin{widetext}
\begin{equation}
\label{eq:CurrMinTURSIJP}
\begin{split}
v_{xy}^*(\newtext{\tau})
&=
\frac{
v_{xy}^{\bar{\rho}}(\newtext{\tau})\left(
\mathfrak{j}
-
\int_{0}^{\infty} e^{-\newtext{\tau}'}\, v_{xy}^{\bar{\rho}}(\newtext{\tau}')\, d\newtext{\tau}'
\right)
}{
\int_{0}^{\infty} e^{-\newtext{\tau}'}\, v_{xy}^{\bar{\rho}}(\newtext{\tau}')\,
\operatorname{arcsinh}^{-1}\!\left(
\frac{
v_{xy}^{\bar{\rho}}(\newtext{\tau}')
}{
\sqrt{\big(u_{xy}^{\bar{\rho}}(\newtext{\tau}')\big)^2-\big(v_{xy}^{\bar{\rho}}(\newtext{\tau}')\big)^2}
}
\right)\, d\newtext{\tau}'
}
\;\operatorname{arcsinh}^{-1}\!\left(\frac{v_{xy}^{\bar{\rho}}(\newtext{\tau})}{\sqrt{\big(u_{xy}^{\bar{\rho}}(\newtext{\tau}')\big)^2-\big(v_{xy}^{\bar{\rho}}(\newtext{\tau}')\big)^2}}\right) + v_{xy}^{\bar{\rho}}(\newtext{\tau})\, .
\end{split}
\end{equation}
\end{widetext}

Regardless of the specific choice for the current $v^*(\newtext{\tau})$, the bound \eqref{eq:CurrRateUppTURSIJP} has a specific meaning: the instantaneous entropy production rate of the SIJP, which we can interpret as a measure of dynamical dissipation, bounds all asymptotic current fluctuations.

\subsection{SIJP Thermodynamic Uncertainty Relation}
\label{subsec:TUR}

We now restrict our analysis to the scalar observable $D_T$ in \er{eq:GenCurrSIJP}, whose most general realisation can be written as
\begin{equation}
\label{eq:GenCurrFlucSIJP}
j = \sum_{x < y} \beta_{xy} \mathfrak{j}_{xy} \, , 
\end{equation}
where $\mathfrak{j}_{xy}$ is defined in \er{eq:CurrRecastTURSIJP0}. We aim at calculating an upper bound of the rate function $I(j)$ as follows:
\begin{equation}
            \label{eq:ScalUppBoundTURSIJP}
            \begin{split}
            I(j) &\leq \\ &\hspace{-0.6cm} \inf_{\substack{v(\newtext{\tau}) \\ j: \text{Eq.\ } \eqref{eq:GenCurrFlucSIJP} }} \sum_{x < y} \int_0^\infty e^{-\newtext{\tau}} \frac{(v_{xy}(\newtext{\tau}) - v^{\bar{\rho}}_{xy}(\newtext{\tau}))^2}{4 (v^{\bar{\rho}}_{xy}(\newtext{\tau}))^2} \sigma^{\bar{\rho}}_{xy}(\newtext{\tau}) \, d\newtext{\tau} \, .
            \end{split}
        \end{equation}

We do so by introducing a specific extended Lagrangian incorporating the constraint \eqref{eq:GenCurrFlucSIJP} via a Lagrangian field. The minimisation is explicit and yields the optimal instantaneous edge current
\begin{equation}
                \label{eq:OptCurrdTURSIJP}
                \begin{split}
                v^*_{xy}(\newtext{\tau}) &= v^{\bar{\rho}}_{xy}(\newtext{\tau}) + \\ 
                &\hspace{-0.5cm} \frac{(j - \bar{j}) \beta_{xy} (v^{\bar{\rho}}_{xy}(\newtext{\tau}))^2}{ \sigma^{\bar{\rho}}_{xy}(\newtext{\tau}) \sum_{x < y} \int_0^\infty e^{-\newtext{\tau}'} \frac{(\beta_{xy} v^{\bar{\rho}}_{xy}(\newtext{\tau}'))^2}{\sigma^{\bar{\rho}}_{xy}(\newtext{\tau}')} \, d\newtext{\tau}'} \, ,
                \end{split}
            \end{equation}
where
\begin{align}
			\bar{j} &\coloneqq \int_0^\infty e^{-\newtext{\tau}} j^{\bar{\rho}}_\newtext{\tau} \, d\newtext{\tau} \\
            j_\newtext{\tau}^{\bar{\rho}} &\coloneqq \sum_{x < y} \beta_{xy} v^{\bar{\rho}}_{xy}(\newtext{\tau}) \, .
        \end{align}
        
Inserting $v^*(\newtext{\tau})$ from \er{eq:OptCurrdTURSIJP} in \er{eq:ScalUppBoundTURSIJP} we get
\begin{equation}
                \label{eq:ScalUppBounddTURSIJP}
                I(j) \leq \frac{(j - \bar{j})^2}{4 \sum_{x < y} \int_0^\infty e^{-\newtext{\tau}'} \frac{(\beta_{xy} v^{\bar{\rho}}_{xy}(\newtext{\tau}'))^2}{\sigma^{\bar{\rho}}_{xy}(\newtext{\tau}')} \, d\newtext{\tau}' } \, .
            \end{equation}
Eventually, by applying the Cauchy--Schwarz relation in \er{eq:CauchySchwarz}, with $a^{-1}_{xy}(\newtext{\tau}) = \sigma^{\bar{\rho}}_{xy}(\newtext{\tau})$ and $b_{xy}(\newtext{\tau}) = \beta_{xy} v^{\bar{\rho}}_{xy}(\newtext{\tau})$ in its saturated form, we obtain
\begin{equation}
            \label{eq:ScalUppBounddFinalTURSIJP}
            I(j) \leq I^\Box(j) \coloneqq\frac{(j - \bar{j})^2}{4} \left( \int_0^\infty e^{-\newtext{\tau}} \frac{(j_\newtext{\tau}^{\bar{\rho}})^2}{\sigma_\newtext{\tau}} \, d\newtext{\tau} \right)^{-1} \, ,
        \end{equation}
where $\sigma_\newtext{\tau}$ is the total instantaneous entropy production rate of the SIJP, given by 
\begin{align}
\label{eq:TotEntrProdSIJP}
\sigma_t &= \sum_{x < y} \sigma^{\bar{\rho}}_{xy}(\newtext{\tau}) \\
&=\sum_{x < y} \log \left( \frac{\bar{\rho}_x(\newtext{\tau}) Q_{xy}(\bar{M}_\newtext{\tau})}{\bar{\rho}_y(\newtext{\tau}) Q_{yx}(\bar{M}_\newtext{\tau})} \right) v^{\bar{\rho}}_{xy}(\newtext{\tau}) \, , 
\end{align}
where the second line can be derived from \er{eq:InstEdgeEntrProdSIJP} using the ``$\arcsinh - \log$'' relation.

Eventully, the SIJP-TUR is obtained by expanding the l.h.s\ of \er{eq:ScalUppBounddFinalTURSIJP} up to second order  locally around $\bar{j}$. This yields
\begin{equation}
            \label{eq:TURSIJPs}
            \frac{\left( \text{var}(j) \right)}{\bar{j}^2} \geq 2\int_0^\infty e^{-\newtext{\tau}} \frac{(j_\newtext{\tau}^{\bar{\rho}})^2}{\sigma_\newtext{\tau}} \, d\newtext{\tau}   \, ,
\end{equation}
as a generalisation to SIJPs of the well-known TUR for continuous-time Markov processes~\cite{barato2015thermodynamic,gingrich2016dissipation}. In analogy to the Markovian case, \er{eq:TURSIJPs} shows that reducing current fluctuations in a SIJP incurs a minimal cost in terms of the instantaneous entropy production $\sigma_\newtext{\tau}$. Notice also that it is not possible to disentangle the instantaneous entropy cost from the instantaneous generalised current. As a consequence, we can interpret the instantaneous entropy production as an additional discount for the SIJP, on top of the exponential one, that is necessary to reach a certain level of precision for the current. \deletetext{Lastly, w}\newtext{W}e \newtext{further} note that in the time-homogeneous case, the \er{eq:TURSIJPs} clearly reduces to the standard TUR of Markov jump processes.

\newtext{The present TUR and entropy production construction is similar in spirit to~\cite{Caprini2019,di-terlizzi2020a-thermodynamic}, but different in implementation. The similarity is that these works also introduce specific auxiliary processes and quantify irreversibility through a comparison of their path weights, or dynamical costs, with respect to the original process. Here, however, the non-Markovianity does not relate to Langevin processes and does not come from coloured noise or an explicit memory kernel, but from the dependence of the jump rates on empirical observables of the past trajectory. The instantaneous cost entering our entropy-production rate is obtained from the level-2.5 description of empirical occupations and fluxes, after introducing an auxiliary time-inhomogeneous Markov generator that makes the atypical SIJP trajectory typical.}


\section{Illustrative Examples}
\label{sec:Ex}

In this Section, we illustrate the general results above with several simple concrete examples of two- and three-state systems. 

We initially focus on a two-state model where some analytical treatment of the fluctuations can be done, although for the exact rate functions showed the solution is approachable only by numerical methods. We will derive so-called level-2 LD functions (LD functions for the state occupation), and compare them to simulations and to the upper bounds given by Markov approximations. Along with this, we will study the LD functions of the unidirectional flux in the system and compare them to simulations and to the SIJP-KUR upper bounds given in \er{eq:IbBoundAlphaTimeSIJP}. 

Then, we discuss three-state examples which have a non-equilibrium stationary state. We will see that even for such simple systems an efficient method to compute numerically the exact rate function for the total current through the system is out of reach at the moment and we leave this as a\deletetext{s}\newtext{n} \deletetext{a}\newtext{open} problem to study in future work. We therefore compare simulations of the process with the SIJP-TUR of \er{eq:ScalUppBounddFinalTURSIJP}, testing the validity of the latter as an upper bound to the true fluctuations.

Finally, in the last example, we go back to a simpler, effectively one-state, stochastic birth process where the specific self-interaction considered leads to two (not one) basins of attractions for the long-time empirical birth rate. The large deviations of such an observable reveals striking dynamical features, \deletetext{including a change of convergence speed and the appearance of non-convex behaviour in the strong fluctuation regime.}\newnewtext{such as a change of convergence speed.}


\subsection{Feedback two-state SIJP}
\label{subsec:twostate}

We consider SIJPs for a two-state system, $\mathcal{S} = \left\lbrace 0,1 \right\rbrace$, with rate matrix
\begin{equation}
Q(A_t) \coloneqq
        \begin{pmatrix}
    -1 - h(L_1(t)) & 1 + h(L_1(t)) \\
    1 & -1
\end{pmatrix} \, .	        
\end{equation}
The specific dynamics depends on the dependence on the empirical distribution through the function $h$. We study two cases, 
\begin{align}
    h^{({\rm L})}(L_1(t)) &= L_1(t) \\
    h^{({\rm E})}(L_1(t)) &= \exp{[\alpha \, L_1(t)]} \, ,
\end{align}
for $\alpha > 0$. This describes a spin whose flip-up rate grows either linearly (L) or exponentially (E) with its past occupation. Either of these forms define a SIJP as in \ers{sec:model}{eq:EmpDepSIJPConfig}, where the empirical observable $A_t = L_1(t)$ is defined by setting $(f_0,f_1) = (0,1)$ and $g=0$ in \er{eq:EmpDepSIJPConfig}.

In both cases, the stationary state is found by explicitly solving \er{eq:StatSIJP}. In the linear case (L), the explicit solution to the stationary distribution is 
\begin{equation}
    \label{eq:TypStateEx1}
    (\pi_0^{({\rm L})}, \pi_1^{({\rm L})}) = \left( \frac{3-\sqrt{5}}{2}, \frac{\sqrt{5} - 1}{2} \right) \, ,
\end{equation}
which shows that the system spends on average $12\%$ less time at state $0$ than at state $1$. In the second case, the stationary state $(\pi_0^{({\rm E})}, \pi_1^{({\rm E})})$ is solution of the following transcendental equation:
\begin{equation}
    \label{eq:TypStateEx2}
    2 + e^{\alpha (1-\pi_0^{({\rm E})})} = \frac{1}{\pi_0^{({\rm E})}} \, ,
\end{equation}
which, depending on $\alpha$, returns a value in the interval $0 < \pi_0^{({\rm E})}<\frac{1}{3}$, where the right boundary is obtained by setting $\alpha = 0$ and the left for $\alpha \rightarrow \infty$. The larger $\alpha$, the less time the system typically spends in state $0$ (for $\alpha = 2$ we have $\pi_0^{({\rm E})} \approx 0.13$).

\subsubsection*{Level-2 large deviations}

In the following, we study the level-2 rate function $I_2(\ell)$ for the occupation distribution at $0$, that is, $L_0(t)$, along with the optimal time-dependent fluxes that lead to its form.

For such a simple two-state system we can simplify notation. In the following, we drop the superscript denoting the specific form of $h$, as the treatment is general, and restore it when deriving specific results. For compactness we also define $\pi \coloneqq \pi_0$, $\rho_\newtext{\tau} \coloneqq \rho_0(\newtext{\tau})$, and \newtext{recall that} $M_\newtext{\tau}$ \newtext{is the the empirical observable $A_t$ over the backward logarithmic time-scale}\deletetext{$\coloneqq M_0(t)$}, and due to normalisation we have $\pi_1 = 1 - \pi$, $\rho_1(\newtext{\tau}) = 1 - \rho_\newtext{\tau}$\deletetext{, and $M_1(t) = 1 - M_\newtext{\tau}$}. Additionally, by conservation of probability, the overall net current is zero and therefore $\eta_{01}(\newtext{\tau}) = \eta_{10}(\newtext{\tau}) \eqqcolon \eta_\newtext{\tau}$. All this helps simplify \er{eq:25FunctionalFinalEquivalent}, which takes the form
    \begin{equation}
        \label{eq:JEx1}
        \begin{split}
        \tilde{\mathcal{I}}[\rho,\eta] &= \int_0^\infty e^{-\newtext{\tau}} \bigg( \rho_\newtext{\tau} (1+h(\newtext{M_\newtext{\tau}})) \times \\ &\hspace{-1cm} \Gamma \left( \frac{\eta_\newtext{\tau}}{\rho_\newtext{\tau} (1+h(\newtext{M_\newtext{\tau}}))} \right) + (1 - \rho_\newtext{\tau}) \Gamma \left( \frac{\eta_\newtext{\tau}}{1-\rho_\newtext{\tau}} \right) \bigg) \, d\newtext{\tau} \, .
        \end{split}
    \end{equation}

The level-2 rate function is obtained by minimising $\tilde{\mathcal{I}}[\rho,\eta]$ over $(\rho_\newtext{\tau},\eta_\newtext{\tau})$ under the constraints given by Eqs.\ \eqref{eq:RareDensity} and \eqref{eq:Ms}. A free minimisation over $\eta_\newtext{\tau}$ yields the following optimal flux
\begin{equation}
    \label{eq:OptFluxEx}
    \eta_\newtext{\tau}^* = \sqrt{\rho_\newtext{\tau} (1-\rho_\newtext{\tau}) (1+h(\newtext{M_\newtext{\tau}}))} \, ,
\end{equation}
and the simplified functional
\begin{equation}
    \label{eq:QuadraticEx1}
    \begin{split}
    \tilde{\mathcal{I}}[\rho,\eta^*] &= 
    \\
    &\hspace{-1cm}\int_0^\infty e^{-\newtext{\tau}} \left( \sqrt{\rho_\newtext{\tau} (1+h(\newtext{M_\newtext{\tau}}))} - \sqrt{1-\rho_\newtext{\tau}} \right)^2 \, d\newtext{\tau} \, .
    \end{split}
\end{equation}

Noticeably, its quadratic form guarantees that there is a unique minimising trajectory $\rho^*_\newtext{\tau}$ leading to the realisation of the fluctuation $L_0(t) = \ell$ in the long-time limit. In the typical state, obtained by equating the last display to $0$, we readily get
\begin{equation}
    \label{eq:NaiveEx1}
    \rho^*_\newtext{\tau} = \newtext{1-M_\newtext{\tau}^*} = \pi \, ,
\end{equation}
along with the corresponding optimal constant flux in \er{eq:OptFluxEx}. 

Minimising \er{eq:QuadraticEx1} over $\rho_\newtext{\tau}$ with the constraints \eqref{eq:Ms} and \eqref{eq:RareDensity} does not lead to an explicit formula for $\rho^*_\newtext{\tau}$ and neither for the level-2 rate function. Yet, a closed form solution for $\rho^*_\newtext{\tau}$ can be found by introducing the extended Lagrangian
\begin{equation}
        \label{eq:LagrangianEx1}
        \begin{split}
        \mathcal{L}[\rho_\newtext{\tau}, \zeta_\newtext{\tau}, \varepsilon] &\coloneqq \\
        &\hspace{-2cm} \int_0^\infty e^{-\newtext{\tau}} \bigg[ \left( \sqrt{\rho_t (1+h(\newtext{M_\newtext{\tau}}))} - \sqrt{1-\rho_\newtext{\tau}} \right)^2 + \\
        &\hspace{-1cm}\zeta_\newtext{\tau} (\dot{M}_\newtext{\tau} - M_\newtext{\tau} + \newtext{1-} \rho_\newtext{\tau} ) + \varepsilon e^{-\newtext{\tau}} \left( \rho_\newtext{\tau} - \pi \right) \bigg] \, d\newtext{\tau} \, ,
        \end{split}
    \end{equation}
where $\zeta_\newtext{\tau} \in \mathbb{R}$ for all $\newtext{\tau} \in \mathbb{R}_+$ and $\varepsilon \in \mathbb{R}$ are the Lagrangian fields enforcing the constraints. In particular, notice that \er{eq:Ms} has been enforced as a differential equation rather than an integral equation, which is more handy.

Minimising variationally \er{eq:LagrangianEx1} and  replacing the Euler--Lagrange equation for $\rho_\newtext{\tau}$ in the costate equation for $M_\newtext{\tau}$ we obtain, after some algebra, the  following system of equations:
\begin{align}
        \label{eq:MsDiffEx1}
            \dot{M}_\newtext{\tau} &= M_\newtext{\tau} - \newtext{1-}\rho_\newtext{\tau} \\
        \label{eq:rhosDiffEx1}
            \dot{\rho}_\newtext{\tau} &= r(M_\newtext{\tau},\rho_\newtext{\tau}) \, ,
\end{align}
where the function $r$ takes the form
\begin{widetext}
\begin{align}
    \label{eq:g1Ex}
    \begin{split}
    r(M_t,\rho_t) &= - \frac{d h(\newtext{M_\tau})}{d M_\newtext{\tau}} \bigg( \frac{\rho_\newtext{\tau}^3 (1 - 2 M_\newtext{\tau}) + \rho_\newtext{\tau}^2 (3 M_\newtext{\tau} - 1 - 2 \sqrt{1+h(\newtext{M_\tau})} \newtext{1-M_\tau} \sqrt{\rho_\newtext{\tau}(1-\rho_\newtext{\tau})})}{1+h(\newtext{M_\tau})} + \\
    &\hspace{7cm}\frac{M_\newtext{\tau} \rho_\newtext{\tau} (2 \sqrt{1+h(\newtext{M_\tau})}\sqrt{\rho_\newtext{\tau}(1-\rho_\newtext{\tau})} - 1)}{1+h(\newtext{M_\tau})} \bigg) \, .
    \end{split}
\end{align}
\end{widetext}

The system of equations is to be solved with the initial condition $M_0 = \newtext{1-}\ell$ and an initial condition for $\rho_0$, which can only be found a posteriori by setting the integral constraint over the solution $\rho_\newtext{\tau}$ to match $\ell$, that is, $\int_0^\infty e^{-\newtext{\tau}} \rho_\newtext{\tau} \, d\newtext{\tau} = \ell$. 

\begin{figure*}[t]
  \centering
  \begin{minipage}{0.49\textwidth}\centering
    \includegraphics[width=\linewidth]{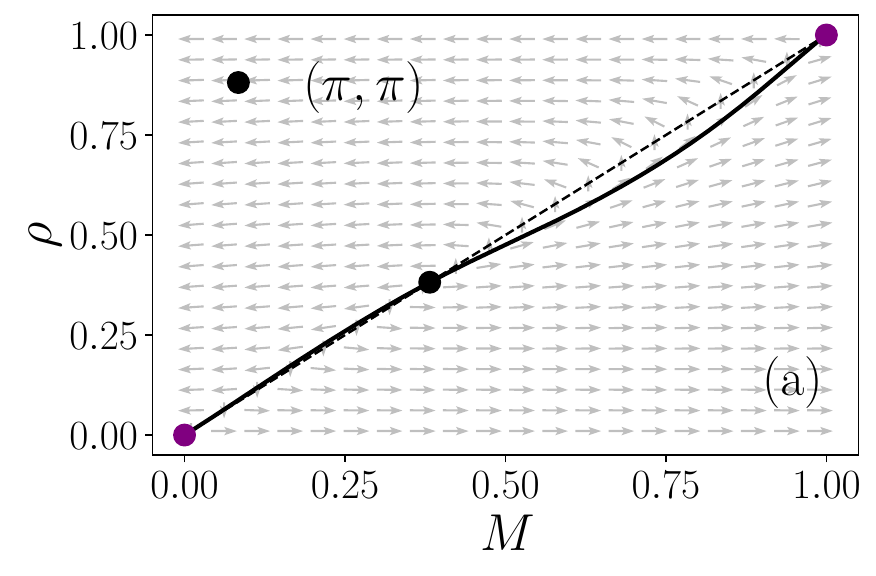}
  \end{minipage}\hfill
  \begin{minipage}{0.49\textwidth}\centering
    \includegraphics[width=\linewidth]{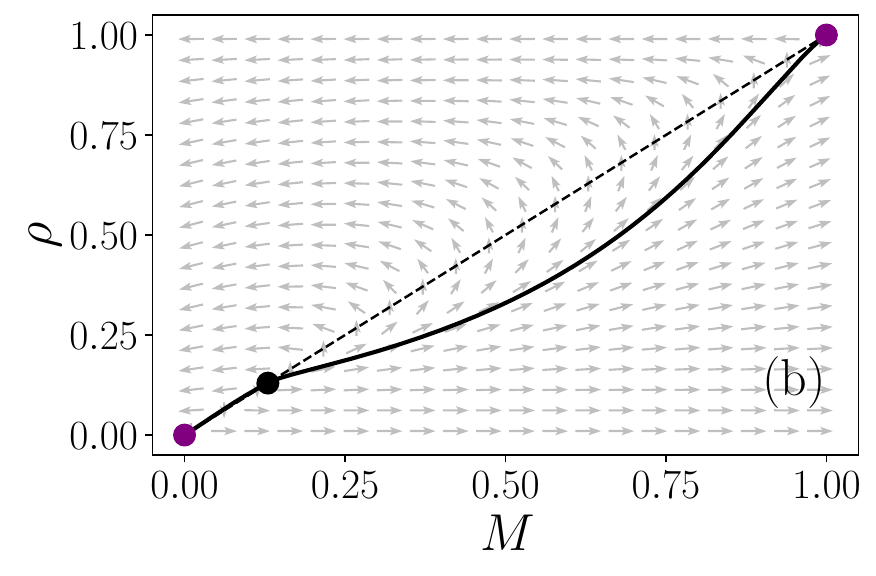}
  \end{minipage}
  \caption{Phase portraits of the optimal trajectories $(\rho_\newtext{\tau},M_\newtext{\tau})$ solving Eqs.~\eqref{eq:MsDiffEx1}–-\eqref{eq:rhosDiffEx1} for the two-state SIJP with (a) linear feedback $h^{({\rm L})}(L_1)=L_1$ and (b) exponential feedback $h^{({\rm E})}(L_1)=\exp(\alpha L_1)$ with $\alpha=2$. Fixed points at $(0,0)$, $(\pi,\pi)$, and $(1,1)$ are shown, with $(0,0)$ and $(1,1)$ acting as saddle points (purple circles). The black solid line denotes the one-dimensional stable manifold connecting the saddle points through $(\pi,\pi)$, along which all admissible optimal trajectories lie. Arrows indicate the direction of increasing reversed time \newtext{$\tau$}.
}
  \label{fig:phasediagram}
\end{figure*}

\begin{figure*}[t]
  \centering
  \begin{minipage}{0.49\textwidth}\centering
    \includegraphics[width=\linewidth]{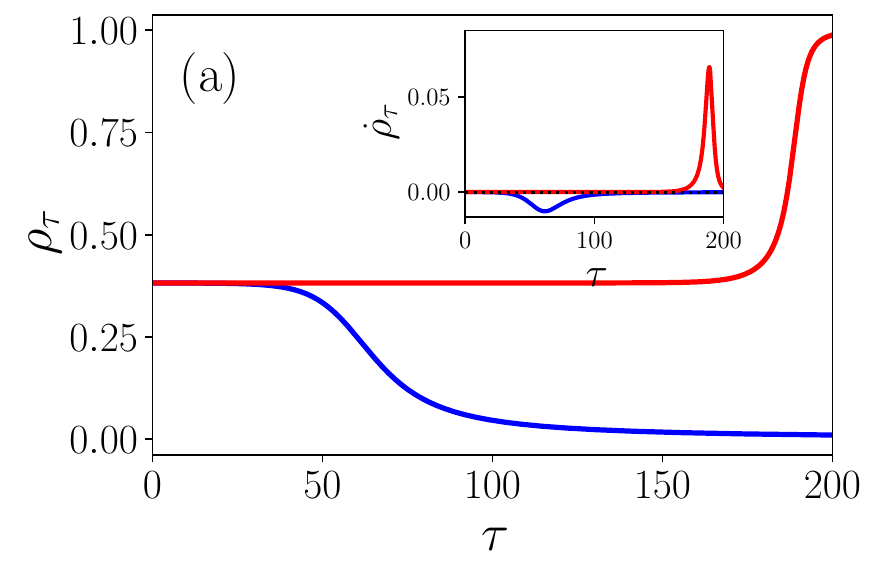}
  \end{minipage}\hfill
  \begin{minipage}{0.49\textwidth}\centering
    \includegraphics[width=\linewidth]{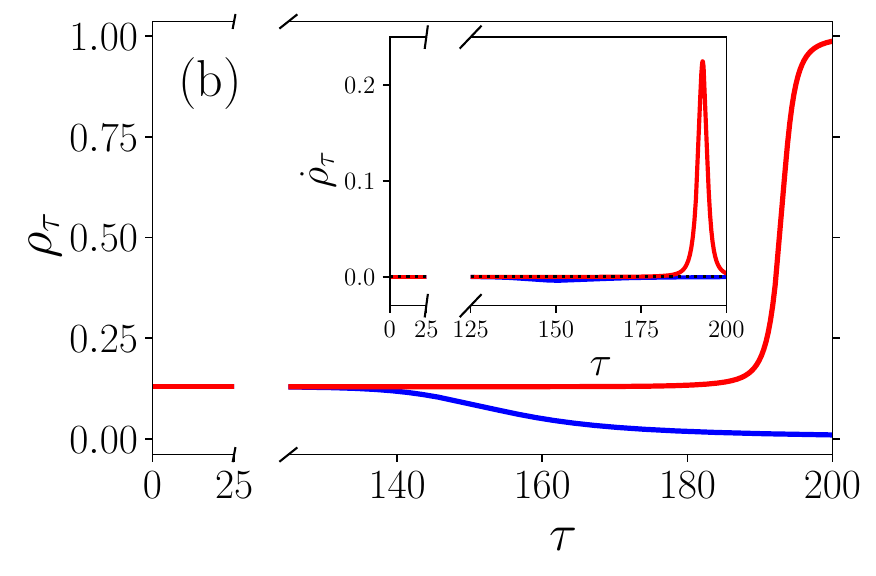}
  \end{minipage}
  \caption{Optimal trajectories $(\rho_\newtext{\tau},M_\newtext{\tau})$ along the stable manifold for the two-state SIJP with (a) linear and (b) exponential feedback. Trajectories are obtained by integrating\newtext{, backward with respect to time $\tau$, }Eqs.~\eqref{eq:MsDiffEx1}–-\eqref{eq:rhosDiffEx1} \deletetext{backward in time}from initial conditions close to the saddle points\newtext{: the red trajectory starting close to $(1,1)$ and the blue trajectory starting close to $(0,0)$}. Each point on the curves corresponds to a distinct initial value $M_0=\newtext{1-}\ell$, generating a fluctuation of the empirical occupation measure $L_0(T)=\ell$ in the (physical) long-time limit. Trajectories converge to $(\pi,\pi)$ as $\newtext{\tau}\to0$.
}
  \label{fig:trajs}
\end{figure*}

For both cases, the system of ordinary differential equations above is highly unstable because all its fixed points, $(\rho, M) \in \left\lbrace(0,0), (\pi,\pi), (1,1) \right\rbrace$, are unstable. The phase diagrams in Fig.\ \ref{fig:phasediagram} show that $(0,0)$ and $(1,1)$ act as saddle points whose one-dimensional stable manifolds intersect at $(\pi,\pi)$. Consequently, all initial conditions $(\rho_0, M_0)$ must lie on this stable manifold (black solid lines). It is therefore sufficient to compute two sets of trajectories, $(\rho_\newtext{\tau}, M_\newtext{\tau})$, with initial conditions chosen near $(\pi,\pi)$ on the stable manifold---one on each side of the maximum corresponding to the two saddle points. Since these trajectories evolve along the stable manifold and asymptotically converge to their respective saddle points, together they span all possible initial conditions. Hence, to study trajectories starting from different initial states, it suffices to vary the \newtext{backward-}time parameter $\newtext{\tau}$ until $(\rho_\newtext{\tau}, M_\newtext{\tau})$ reaches the desired condition. We plot the two sets of trajectories for both systems under analysis in Fig.\ \ref{fig:trajs}. [Numerically, to deal with instability, we study the system of differential equations\deletetext{backward, rather than forward, in time. Hence, we} \newtext{starting} in the vicinity of the saddle points and follow trajectories as they converge to $(\pi,\pi)$.]

\begin{figure*}[t]
  \centering
  \begin{minipage}{0.49\textwidth}\centering
    \includegraphics[width=\linewidth]{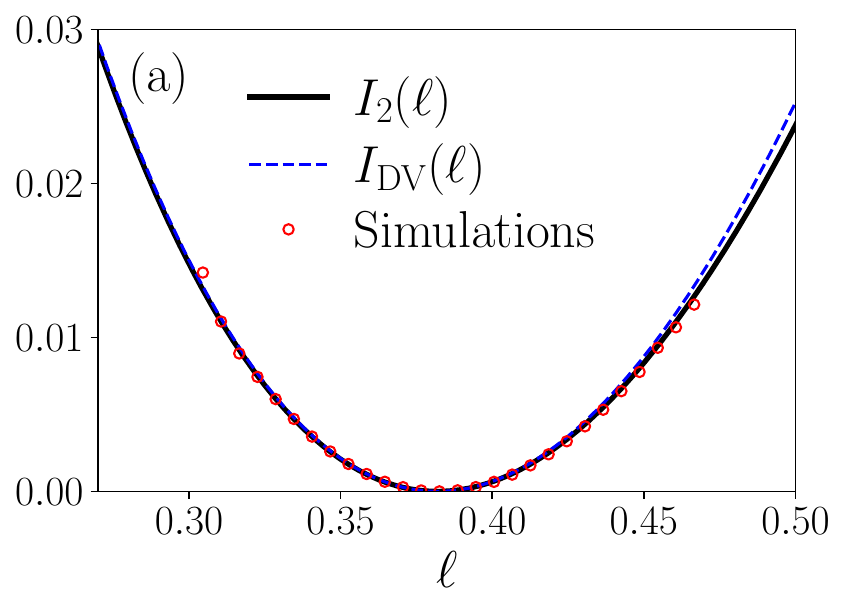}
  \end{minipage}\hfill
  \begin{minipage}{0.49\textwidth}\centering
    \includegraphics[width=\linewidth]{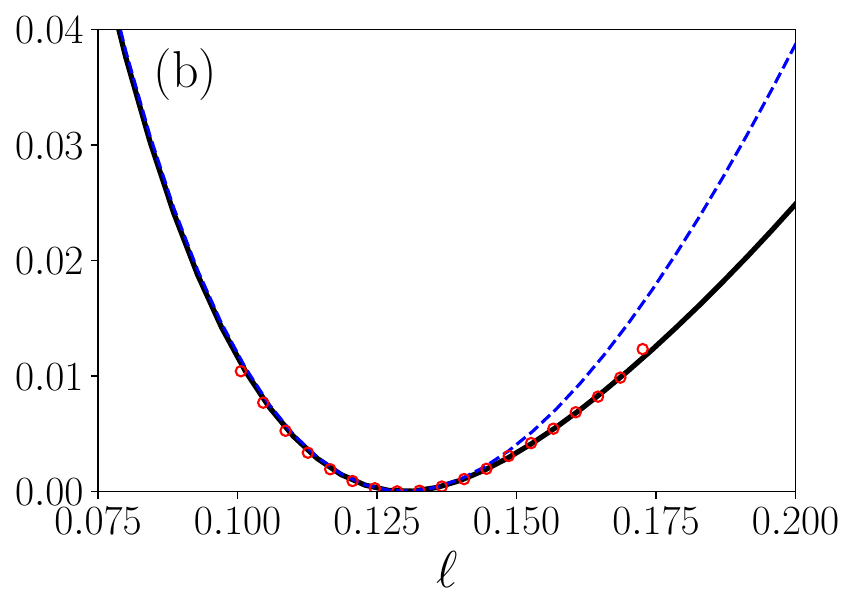}
  \end{minipage}
  \caption{Level-2 large deviation rate function $I_2(\ell)$ of the empirical occupation measure $L_0(T)$ for the two-state SIJP with (a) linear and (b) exponential feedback. Solid black lines show the exact rate function obtained by evaluating Eq.~\eqref{eq:QuadraticEx1} along the optimal trajectories $(\rho_\newtext{\tau}^*,M_\newtext{\tau}^*)$. Symbols correspond to Monte Carlo simulations. The dashed curve shows the Markov (Donsker–Varadhan) upper bound $I_{\mathrm{DV}}(\ell)$ obtained from the constant-density Ansatz $\rho_\newtext{\tau}=\ell$. Deviations from the Markov bound become pronounced for large fluctuations away from $\pi$.
}
  \label{fig:ratelev2}
\end{figure*}

In Fig.\ \ref{fig:ratelev2}, we compare the level-2 rate function $I_2(\ell)$ for the SIJP, obtained by plugging $(\rho_\newtext{\tau}^*,M_\newtext{\tau}^*)$ for each value $L_0(\newtext{\tau}) = \ell$ into \er{eq:QuadraticEx1} with Monte-Carlo simulations---whose detailed explanations is deferred to Appendix \ref{app:sim}---and with the Markov upper bound $I_{\text{DV}}(\ell)$. The latter is provided by the na\"{i}ve (constant) suboptimal solution 
\begin{equation}
    \label{eq:NaiveDVEx1}
    \rho_\newtext{\tau}^* = \newtext{1-}M_\newtext{\tau}^* = \ell \, ,
\end{equation}
which would be the correct ansatz, leading to the Donsker--Varadhan rate function, if the process were Markov and time-homogeneous.

As expected, in a neighbourhood of $\pi$, where the rate functions attain their minimum and vanish, the Markov upper bound closely matches the exact rate function. However, for fluctuations further away from $\pi$, $I_2(\ell)$ departs from $I_{\mathrm{DV}}(\ell)$. This discrepancy arises because LDs from the mean require increasingly abrupt changes in the optimal trajectories $(\rho_\newtext{\tau}, M_\newtext{\tau})$, as illustrated in Fig.\ \ref{fig:trajs}. In particular, trajectories producing fluctuations with $\ell>\pi$ evolve more rapidly—exhibiting larger slopes in absolute value—so that significant deviations from $\pi$ are reached at earlier times, when the exponential discount $e^{-\newtext{\tau}}$ is still appreciable. As a result, the right tails of $I_2(\ell)$ and $I_{\mathrm{DV}}(\ell)$ differ markedly, whereas for $\ell < \pi$ the left tails do not show relevant discrepancies.

\subsubsection*{Flux large deviations and the SIJP-KUR}

In the following, we carry out a similar analysis to derive the unidirectional flux rate function $I(b)$, where $b$ is an instance of the generalised observable $B_T = \Phi_{01}(T)$ in \eqref{eq:GenFluxSIJP}. The minimisation over $\tilde{\mathcal{I}}[\rho,\eta]$ in \er{eq:JEx1} requires different constraints for $(\rho_\newtext{\tau},\eta_\newtext{\tau})$. First, we must impose \er{eq:FluctbSIJP} where $\phi_{01} = \phi_{10}$ is fixed by \er{eq:RareFlux}. Minimising over $\eta$ yields
\begin{equation}
    \label{eq:OptFluxRateFluxEx1}
    \eta_\newtext{\tau}^* = \sqrt{\rho_t (1-\rho_t) (1 + h(\newtext{M_\newtext{\tau}}))} \kappa \, ,
\end{equation}
as the optimal flux, where $\kappa$ is the Lagrange parameter fixing the constraint \eqref{eq:FluctbSIJP} playing a useful re-parametrisation of the flux $B_T = b$. 

\begin{figure*}[t]
  \centering
  \begin{minipage}{0.49\textwidth}\centering
    \includegraphics[width=\linewidth]{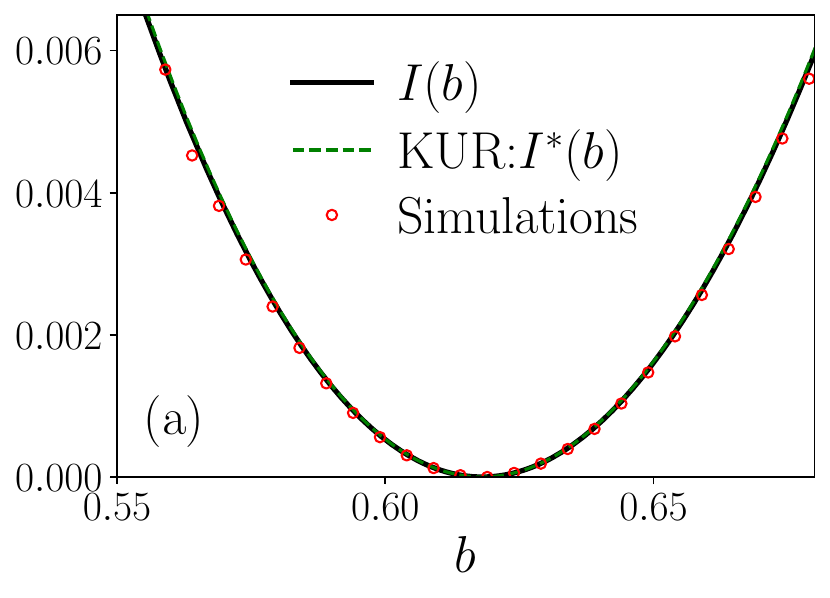}
  \end{minipage}\hfill
  \begin{minipage}{0.49\textwidth}\centering
    \includegraphics[width=\linewidth]{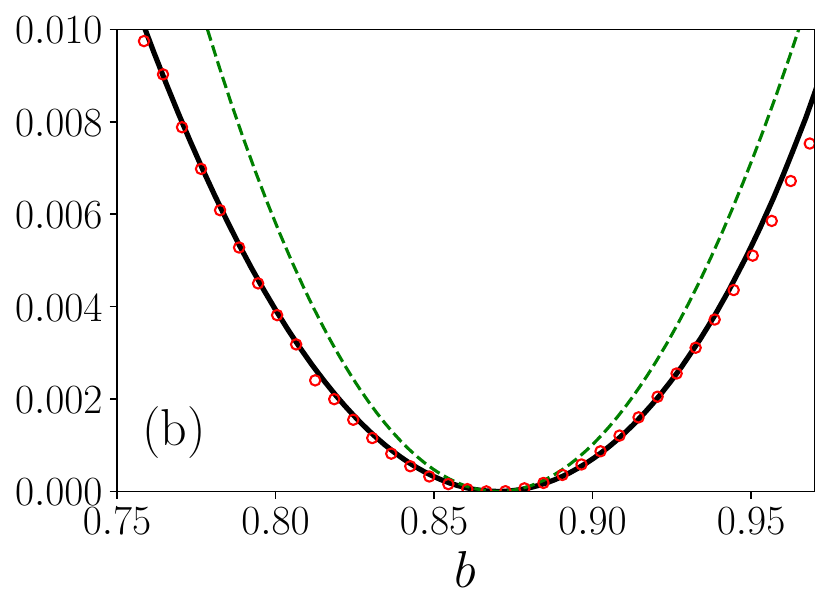}
  \end{minipage}
  \caption{Large deviation rate function $I(b)$ of the unidirectional flux $B_T=\Phi_{01}(T)$ for the two-state SIJP with (a) linear and (b) exponential feedback. Solid black lines denote the exact rate functions obtained by minimising $\tilde{\mathcal{I}}[\rho,\eta]$ under the flux constraint. Symbols correspond to Monte Carlo simulations. The dashed curve shows the SIJP-KUR bound given by Eq.~\eqref{eq:IbBoundSIJP}, which correctly captures typical fluctuations and provides an upper bound on rare events.
}
  \label{fig:ratefluxex1}
\end{figure*}

The next step is to minimise the resulting functional $\tilde{\mathcal{I}}[\rho,\eta^*]$ with respect to $\rho_\newtext{\tau}$ with the constraint given by \er{eq:Ms} (with the specific $f$ and $g$ introduced at the beginning of this Section). Yet, there is no fixed initial condition as we do not a-priori know the optimal density that leads to a specific fluctuation of the flux. To circumvent this obstacle, we consider all possible densities---as done above to derive the level-2 rate function---and by swiping over them keep only the trajectories $(\rho_\newtext{\tau}^*,M_\newtext{\tau}^*)$ that minimise the large deviation functional for a specific flux $b$. This procedure returns the rate functions represented as solid black lines in Fig.\ \ref{fig:ratefluxex1}.

These exact rate functions are matched at the level of typical fluctuations by Monte-Carlo simulations and are upper bounded by the SIJP-KUR function in \er{eq:IbBoundSIJP}.


\subsection{\deletetext{Feedback}\newtext{Occuaption-self-induced rotatory} three-state SIJP}

\subsubsection*{Current large deviations and the SIJP-TUR}


We consider the three-state SIJP, $\mathcal{S}= \left\lbrace 0,1,2 \right\rbrace$, with rate matrix
\begin{equation}
Q(A_t) \coloneqq
        \begin{pmatrix}
    -1 & h(L_1(t)) & 1 - h(L_1(t)) \\
    1 - h(L_1(t)) & -1 & h(L_1(t)) \\
    h(L_1(t)) & 1-h(L_1(t)) & -1 \\
\end{pmatrix} \, ,	        
\end{equation}
where 
\begin{align}
    \label{eq:TanhFeedback}
    h(L_1(t)) = \frac{1}{2} \left[ 1 + \frac{\tanh \left[\alpha (L_1(t) - 1/2) \right]}{\tanh \left[ \alpha/2 \right]}   \right] \, ,
\end{align}
with $\alpha > 0$. This describes a system with a self-induced non-linear push and cyclic transitions. The dynamics tends to accelerate when state $1$ is visited frequently. Yet, the more the particle lingers in a region, the harder it is to sustain a current. Similarly to the previous Subsection, the empirical observable $A_t = L_1(t)$ is obtained by setting $(f_0,f_1\newtext{,f_2}) = (0,1\newtext{,0})$ and $g=0$ in \er{eq:EmpDepSIJPConfig}. Given the inherent rotational symmetry of the system, the stationary distribution solution of \er{eq:StatSIJP} is uniform over $\mathcal{S}$. 

We now focus on the total current observable $J_T$ in \er{eq:GenCurrSIJP} and aim at studying its large deviation function. A priori, given the self-interacting nature of the process, it is not even obvious what the typical behaviour of $J_T$ is in the long-time limit as time-dependent dynamics might lead to deviations from the stationary current of the Markov generator $Q(\bar{a})$. A posteriori, however, after analysing the time-dependent system current averaged over a large ensemble of trajectories, we conclude that the averaged SIJP current is constant over time and, therefore
\begin{equation}
    \label{eq:TypCurrThreeState1}
    \bar{j} = 3 \mathfrak{j} = 3 \left(  (\pi \circ Q(\bar{a}))_{01} - (\pi \circ Q(\bar{a}))_{10} \right) \approx-0.42 \, ,
\end{equation}
where thanks to the rotational symmetry typical edge currents are all equivalent.

\begin{figure*}[t]
  \centering
  \begin{minipage}{0.49\textwidth}\centering
    \includegraphics[width=\linewidth]{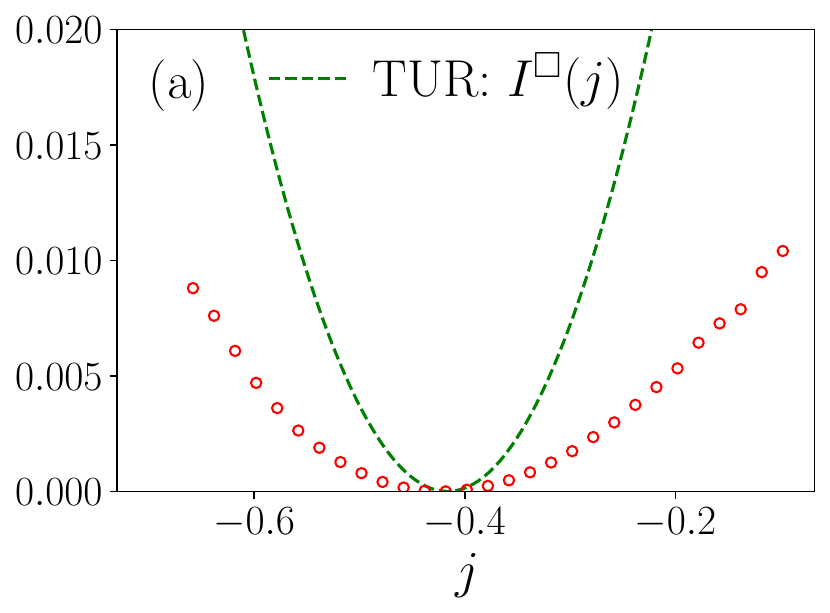}
  \end{minipage}\hfill
  \begin{minipage}{0.49\textwidth}\centering
    \includegraphics[width=\linewidth]{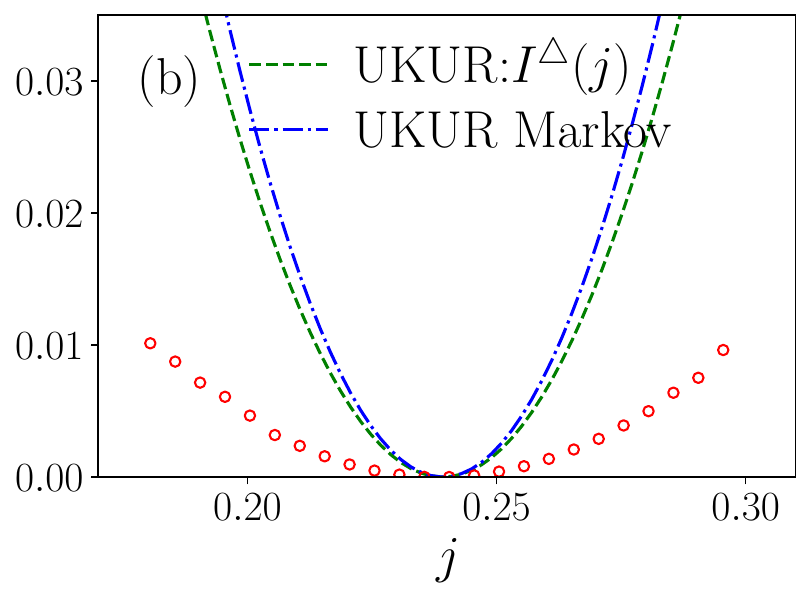}
  \end{minipage}
  \caption{Large deviation behaviour of the total current $J_T$ in three-state SIJPs. (a) Rate function of the total current for the SIJP with state-dependent feedback $h(L_1)$ defined in \er{eq:TanhFeedback}, obtained from Monte Carlo simulations (symbols) and compared with the SIJP-TUR upper bound $I^\Box(j)$ given by Eq.~\eqref{eq:ScalUppBounddFinalTURSIJP} (solid line). (b) Current fluctuations for the SIJP with jump-dependent feedback $A_t=J_t$, compared with the SIJP-UKUR bound $I^\triangle(j)$ in Eq.~\eqref{eq:UKURBoundSimplified3State} (solid line). The dashed line shows the corresponding Markov UKUR for comparison.
}
  \label{fig:rate3state}
\end{figure*}

This scenario is too complicated to attempt analytically the variational minimisation in \er{eq:CurrRateSimplTURSIJP1} and therefore we can only accumulate statistics from simulating the process for a long enough time and compare these simulations to the SIJP-TUR rate function bound $I^\Box(j)$ in \er{eq:ScalUppBounddFinalTURSIJP} (see Fig.\ \ref{fig:rate3state}(a)). We remark that, for this specific system, the SIJP-TUR bound reduces to the well-known Markov TUR bound as $j_\newtext{\tau}^{\bar{\rho}} = \bar{j}$ and $\sigma_\newtext{\tau} = \sigma$, where
\begin{equation}
    \label{eq:MarkovEntrpProd}
    \sigma = \sum_{(x,y) \in \E} (\pi \circ Q(\bar{a}))_{xy} \log \left( \frac{(\pi \circ Q(\bar{a}))_{xy}}{(\pi \circ Q(\bar{a}))_{yx}} \right) \approx 0.37 \, ,
\end{equation}
is the well-known average rate of entropy production for the Markov jump process with generator $Q(\bar{a})$ in the stationary state.

\subsection{\deletetext{Feedback}\newtext{Current-self-induced rotatory} three-state SIJP}

\subsubsection*{Current large deviations and the SIJP-UKUR}

We consider another three-state SIJP, $\mathcal{S}= \left\lbrace 0,1,2 \right\rbrace$, with rate matrix
\begin{equation}
Q(A_t) \coloneqq
        \begin{pmatrix}
    -1 & e^{-\alpha J_t} & 1 - e^{-\alpha J_t} \\
    1 - e^{-\alpha J_t} & -1 & e^{-\alpha J_t} \\
    e^{-\alpha J_t} & 1-e^{-\alpha J_t} & -1 \\
\end{pmatrix} \, ,	        
\end{equation}
with $\alpha > 0$. This again describes a cyclic system. The self-induced exponential drag is enhanced by having a higher current through the system. The empirical observable $A_t = J_t$ is obtained by setting $f=0$ and $g_{01}=g_{12}=g_{20}=1=-g_{10}=-g_{21}=-g_{02}$ in \er{eq:EmpDepSIJPConfig}. Even in this case, the inherent rotational symmetry of the system guarantees a uniform stationary distribution over $\mathcal{S}$. 

We now study the LDs of the total current observable $J_T$ in \eqref{eq:GenCurrSIJP}. Similarly to the previous example, the averaged SIJP current as well as edge currents are constant over time and, therefore,
\begin{equation}
    \label{eq:TypCurrThreeState1}
    \bar{j} = 3 \mathfrak{j} = 3 \left(  (\pi \circ Q(\bar{a}))_{01} - (\pi \circ Q(\bar{a}))_{10} \right) \approx 0.24 \, .
\end{equation}

In Fig.\ \ref{fig:rate3state}(b) we compare the accumulated statistics from simulating the process for a long enough time with the SIJP-UKUR rate function bound $I^\triangle(j)$ in \er{eq:FinalUKURRateBound}. Due to the time-independent behaviour of the average current through the system, the SIJP-UKUR bound simplifies to 
\begin{equation}
    \label{eq:UKURBoundSimplified3State}
    \begin{split}
    I^{\triangle}(j) &= \frac{(j - \bar{j})^2}{2 \bar{j}^2} \frac{e^{-\alpha \bar{j}}}{1-e^{-\alpha \bar{j}}} \times \\
    &\hspace{0.5cm}\left( (3 + \bar{j}) (1 - e^{-\alpha \bar{j}}) - \bar{j} e^{-\alpha \bar{j}} \right)^2 \, ,
    \end{split}
\end{equation}
which is however different from, and improves, the time-homogeneous Markov UKUR (blue dashed line in Fig.\ \ref{fig:rate3state}(b)). This is because in \er{eq:FinalUKURRateBound} a derivative with respect to the self-interaction term appears.

\subsection{Self-induced bistable birth process}

\deletetext{We now consider a self-interacting birth process.}
\newnewtext{We now consider a self-interacting birth process. A birth process is a counting process $N_t$ that increases by one at each birth event, with its dynamics specified by the birth rate $r_t$. Here, this rate depends on the empirical birth activity itself and is given by}
\begin{equation}
    \label{eq:BirthRate}
    r_t = K_t - \frac{(K_t - a)(K_t-b)(K_t-c)}{(K_t^2+p^2)} \, ,
\end{equation}
for $a<b<c \in \mathbb{R}^+$, $p \in \mathbb{R}$, and
\begin{equation}
    \label{eq:EmpiricalRate}
    K_t = \frac{N_t}{t} \, ,
\end{equation}
the time-normalised number of births $N_t$ up to time $t$.

\deletetext{Equivalently, the birth process can be interpreted as a two-state SIJP whereby hops between the two equivalent states occur at rate \eqref{eq:BirthRate}. In this case, Eq.\ \eqref{eq:EmpiricalRate} can be reinterpreted as the empirical total activity of the process, i.e.,}
\newnewtext{Equivalently, the same counting process can be represented as a two-state SIJP with two equivalent states, $0$ and $1$, where each jump $0\leftrightarrow1$ counts as one birth event and both transition rates are given by Eq.\ \eqref{eq:BirthRate}. Since the two states are equivalent, the relevant observable is the total number of jumps rather than the occupation of either state. In this representation, Eq.\ \eqref{eq:EmpiricalRate} is therefore the empirical total activity of the process, i.e.,}
\begin{equation}
    \label{eq:EmpActivity}
    K_t = \Phi_{01}(t) + \Phi_{10}(t) \, ,
\end{equation}
where $\Phi_{xy}(t)$ is the empirical flux defined in Eq.\ \eqref{eq:EmpFluxSIJP}.

The specific form of the rate $r_t$ guarantees bistable behaviour for the observable $K_t$ itself, while the quadratic denominator ensures that the rate remains bounded. This is shown via stochastic approximation by considering the limiting ODE for the empirical activity. The ODE is
\begin{equation}
    \label{eq:LimitingODE}
    \frac{dK_t}{dt} = - \frac{(K_t - a)(K_t-b)(K_t-c)}{t(K_t^2+p^2)} \, . 
\end{equation}
The equation above can also be obtained from the Doob--Meyer decomposition by dropping the martingale contribution, which can be shown to be asymptotically negligible.

\begin{figure}[h!]
    \centering
    \includegraphics[width=1.\linewidth]{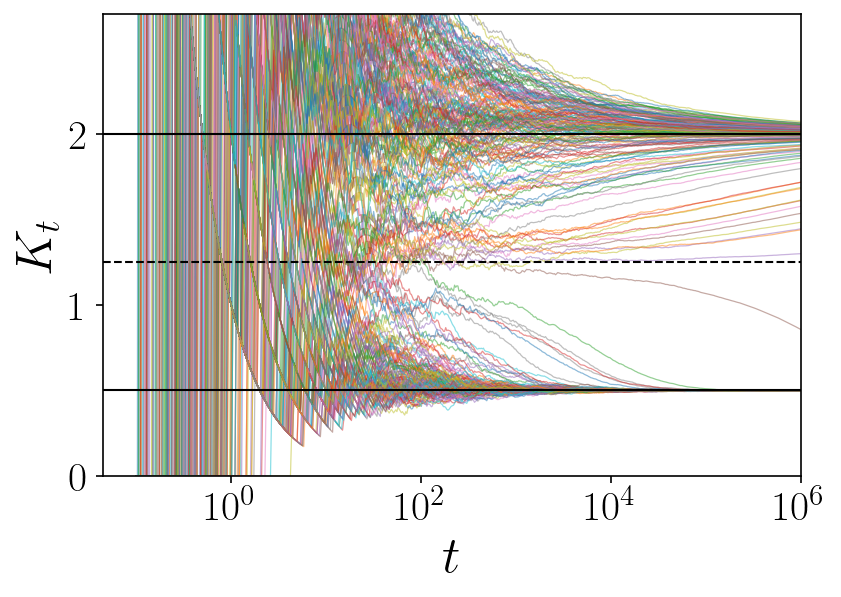}
    \caption{\newtext{Empirical activity trajectory simulations for the birth process with self-interacting birth rate in Eq.\ \eqref{eq:BirthRate}. Trajectories settle at the two stable fixed points $a=0.5$ and $c=2$. The unstable fixed point $b=1.25$ lies between them and dynamically splits trajectories towards the two basins of attraction. Noticeably, trajectories crossing the unstable fixed point settle at a lower ``speed'' than those starting outside the interval. Simulations for $p=0.7$.}}
    \label{fig:SimBirth}
\end{figure}

\deletetext{For simplicity, we move to the logarithmic time scale $u = \log t$, and Eq.\ \eqref{eq:LimitingODE} becomes}
\newnewtext{Using the logarithmic time variable introduced earlier, $s=\log(t/t_0)$ with $t_0=1$, Eq.\ \eqref{eq:LimitingODE} becomes}
\begin{equation}
    \label{eq:LimitingODELog}
    \deletetext{\frac{d K_u}{du} = - \frac{(K_u - a)(K_u-b)(K_u-c)}{(K_u^2+p^2)}}
    \newnewtext{\frac{d K_s}{ds} = - \frac{(K_s - a)(K_s-b)(K_s-c)}{(K_s^2+p^2)}} \, ,
\end{equation}
\deletetext{where, with a slight abuse of notation, we define $K_u \coloneqq K_{e^u}$.}
\newnewtext{where, with a slight abuse of notation, we define $K_s \coloneqq K_{t_0e^s}=K_{e^s}$.}
There are three fixed points for this ODE: $k_\infty = a$, $b$, and $c$. A simple linear stability analysis reveals that $a$ and $c$ are stable, whereas $b$ is unstable. Numerical simulations of the process are reported in Fig.\ \ref{fig:SimBirth} and confirm this analysis. 

To study the large deviations of the empirical activity in Eq.\ \eqref{eq:EmpActivity}, we start from the functional \eqref{eq:25FunctionalFinalEquivalent}. We denote $\rho_\tau \coloneqq \rho_0(\tau)$. Minimisation with respect to $\rho_\tau$, with the normalisation constraint, yields the optimal occupation
\begin{equation}
    \label{eq:OptOccBirth}
    \rho_\tau = \frac{\eta_{01}(\tau)}{\eta_{01}(\tau) + \eta_{10}(\tau)} \, ,
\end{equation}
where $\eta_{xy}(\tau)$ is the instantaneous flux from $x$ to $y$ in the backward rescaled time. Plugging Eq.\ \eqref{eq:OptOccBirth} back into the functional yields 
\begin{equation}
    \label{eq:FunctBirth}
    \begin{split}
        \tilde{\mathcal{I}}[\eta] &= \int_0^\infty e^{-\newtext{\tau}} \bigg( \eta_{01}(\tau) \log \frac{\eta_{01}(\tau)}{\rho_\tau Q(M_\tau)} \\  
        &\hspace{0.5cm}+\eta_{10}(\tau) \log \frac{\eta_{10}(\tau)}{(1-\rho_\tau) Q(M_\tau)} - \eta_{01}(\tau) - \eta_{10}(\tau) \\ &\hspace{0.5cm}+ Q(M_\tau)
        \bigg) \, d\newtext{\tau} \, .
    \end{split}
\end{equation}
This can be further minimised with respect to $\eta_{01}(\tau)$ and $\eta_{10}(\tau)$, together with the two constraints $\dot{M}_\tau = M_{\tau} - \eta_{01}(\tau) - \eta_{10}(\tau)$ and
\begin{equation}
    \label{eq:AtypicalActivityBirth}
    k = \int_0^\infty e^{- \tau} (\eta_{01}(\tau) + \eta_{10}(\tau)) d \tau \, ,
\end{equation}
which represents the long-time atypical value of the empirical activity in the backward rescaled time frame.

\begin{figure*}[t]
  \centering
  \begin{minipage}{0.49\textwidth}\centering
    \includegraphics[width=\linewidth]{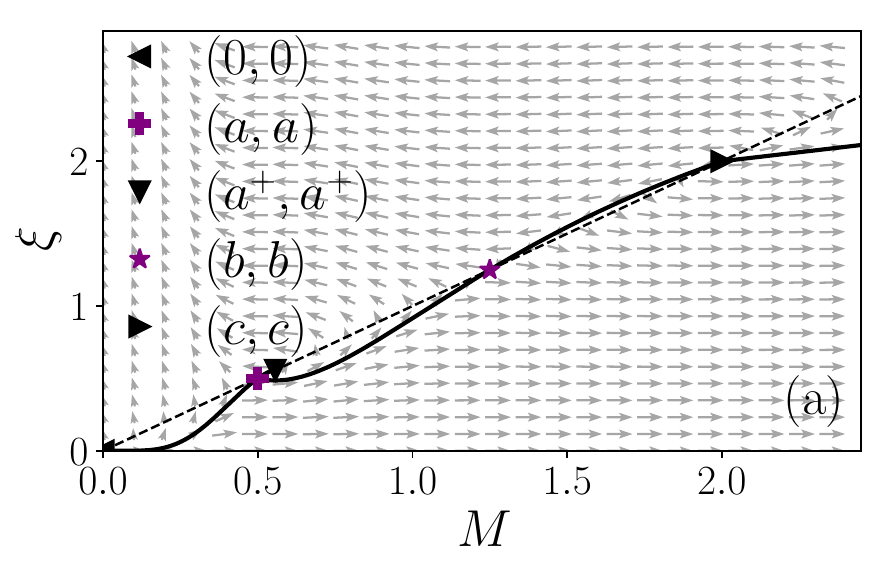}
  \end{minipage}\hfill
  \begin{minipage}{0.49\textwidth}\centering
    \includegraphics[width=\linewidth]{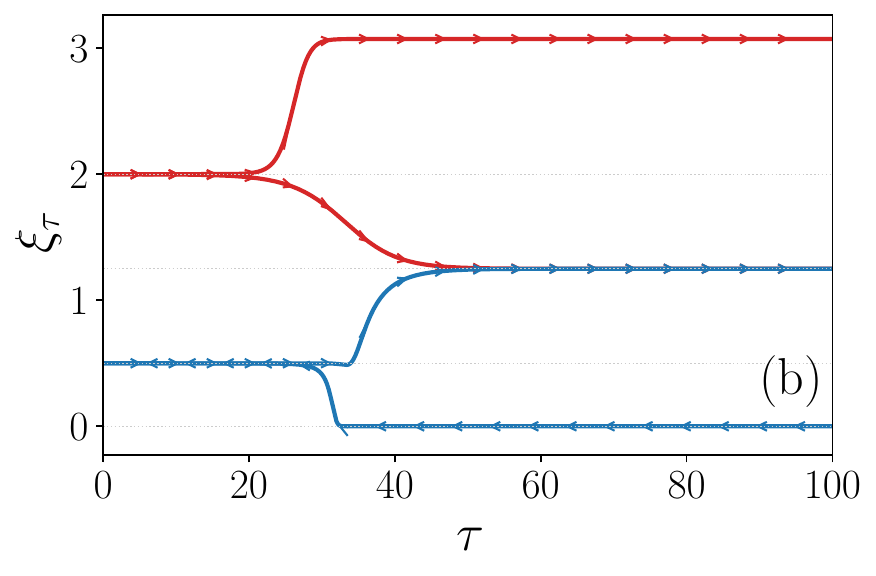}
  \end{minipage}
  \caption{\newtext{(a) Phase portrait of the optimal trajectories \deletetext{$(s_\newtext{\tau},M_\newtext{\tau})$}
\newnewtext{$(\xi_\tau,M_\newtext{\tau})$} solving Eqs.~\eqref{eq:MsDiffBirth}--\eqref{eq:sDiffBirth} for the self-interacting birth process with $a=0.5$, $b=1.25$, $c=2$, and $p=0.7$. Fixed points of the $2$d dynamics are marked with black and purple symbols. The black solid line denotes the one-dimensional stable manifold connecting the saddle points $(a,a)$ and $(b,b)$ with the other unstable points $(0,0)$, $(a^+,a^+)$, and $(c,c)$, along which all admissible optimal trajectories lie. Arrows indicate the direction of increasing reversed time $\tau$. (b) Optimal trajectories \deletetext{$(s_\newtext{\tau},M_\newtext{\tau})$}
\newnewtext{$(\xi_\tau,M_\newtext{\tau})$} along the stable manifold. Trajectories are obtained by integrating Eqs.~\eqref{eq:MsDiffBirth}--\eqref{eq:sDiffBirth}: the red trajectories connect to the fixed point $(c,c)$, while the blue trajectory connects to $(a,a)$. Each point on the curves corresponds to a distinct initial value $M_0=k$, generating a fluctuation of the empirical total activity $K_T=k$ in the physical long-time limit.}}
  \label{fig:Phase+TrajBirth}
\end{figure*}

This yields the following two-dimensional system of ODEs:
\begin{align}
        \label{eq:MsDiffBirth}
            \dot{M}_\newtext{\tau} &= M_\newtext{\tau} - \deletetext{s_\tau}\newnewtext{\xi_\tau} \\
        \label{eq:sDiffBirth}
            \deletetext{\dot{s}_\newtext{\tau}}
            \newnewtext{\dot{\xi}_\tau}
            &= \deletetext{s_\tau}\newnewtext{\xi_\tau} Q'(M_\tau)
            \left[ \frac{M_\tau - Q(M_\tau)}{Q(M_\tau)} \right] \, ,
\end{align}
where \deletetext{$s_\tau = \eta_{01}(\tau) + \eta_{10}(\tau)$.}
\newnewtext{$\xi_\tau = \eta_{01}(\tau) + \eta_{10}(\tau)$ is the total instantaneous flux.} The system must be solved with the initial condition $M_0 = k$ and the terminal condition $\lim_{\tau \to \infty} e^{-\tau} M_\tau = 0$. Both conditions follow from Eq.\ \eqref{eq:AtypicalActivityBirth} using Eq.\ \eqref{eq:MsDiffBirth}. We plot the phase portrait and the trajectory solutions of the ODE system in Fig.\ \ref{fig:Phase+TrajBirth}(a) and (b). 

\begin{figure}[h!]
    \centering
    \includegraphics[width=1.\linewidth]{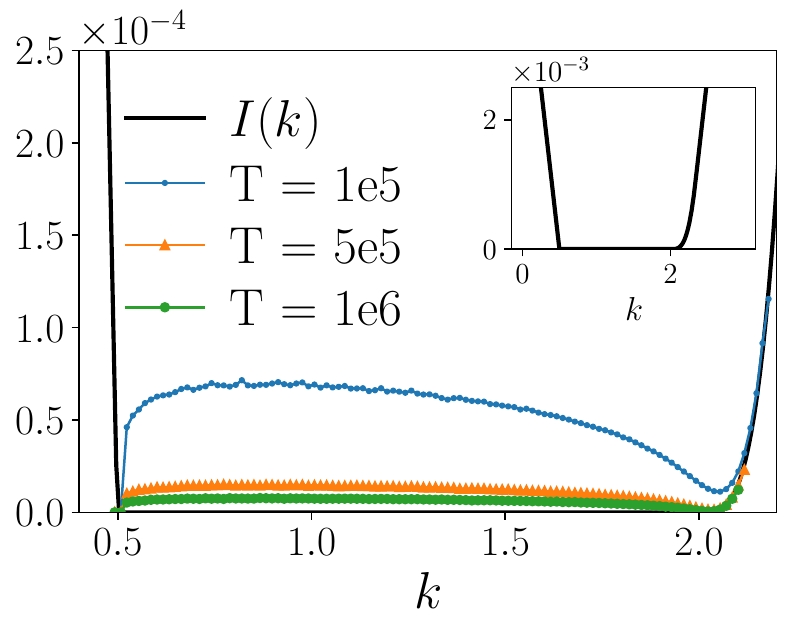}
    \caption{
    \newtext{Large deviation rate function $I(k)$ of the total activity $K_T$ in Eq.\ \eqref{eq:EmpActivity} for the self-interacting birth process. The solid black line denotes the exact rate function obtained by minimising $\tilde{\mathcal{I}}[\rho,\eta]$ under the activity constraint. Symbols correspond to Monte Carlo simulations.}
    }
    \label{fig:RateBirth}
\end{figure}

Finally, in Fig.\ \ref{fig:RateBirth} we plot the rate function of the observable total activity $K_t$ in Eq.\ \eqref{eq:EmpActivity}. The set of zeros of the rate function is not finite and includes all points between $k_\infty = a$ and $k_\infty=c$. Simulations confirm this picture, suggesting that the speed of large deviations may differ for trajectories settling into one of the two stable fixed points from the inner side of the interval compared with trajectories approaching from outside, as Fig.\ \ref{fig:SimBirth} initially suggested. \deletetext{Additionally, as we can notice from the inset in Fig.\ \ref{fig:RateBirth} large deviations are not convex for $k > 2$. The implications of this fact are left for future work.}

\section{Conclusion}
\label{sec:Conclusion}

In this work we have developed a general framework to characterise fluctuations in self-interacting jump processes (SIJPs), a broad class of non-Markovian stochastic dynamics in which transition rates depend on empirical observables of the process itself. Using an exponential tilting construction, we derived the level-2.5 large deviation principle governing the joint fluctuations of the empirical occupation measure and empirical flux for SIJPs with general functional dependence on these quantities. The resulting rate functional reveals a clear separation between fast microscopic dynamics and slow, memory-driven evolution, encoded through an exponential temporal discount. Building on this structure, we obtained variational bounds on flux and current fluctuations, leading to kinetic and thermodynamic uncertainty relations that extend classical Markovian results to non-Markovian settings. Through low-dimensional examples, we demonstrated how self-interaction and time-dependent feedback modify fluctuation behaviour and assessed the tightness of the derived bounds against exact or numerical large deviation results.

Several open questions emerge from this analysis. From a theoretical perspective, the level-2.5 rate function derived here relies on assumptions underpinning the tilting procedure and time-scale separation, which should be examined on a case-by-case basis, particularly for strongly \deletetext{nonlinear or} non-smooth feedbacks. While rigorous large deviation results are available for specific classes of SIJPs, notably those with affine dependence on the empirical measure, extending such results to more general rate maps remains an open challenge. In addition, the possible nonconvexity of the rate function after contraction\deletetext{, which was shown in the bistable example discussed above,} deserves further investigation, as it may signal richer fluctuation mechanisms such as multiple optimal trajectories or dynamical phase transitions. On the level of uncertainty relations, it remains to be understood under which conditions kinetic and thermodynamic bounds are tight, and whether systematically sharper bounds can be obtained by exploiting time-dependent or state-dependent variational Ans\"{a}tze.

Looking ahead, the framework developed here opens several directions for future research. Extending rigorous large deviation techniques to broader classes of self-interacting dynamics would strengthen the mathematical foundations of fluctuation theory for non-Markovian systems. \newtext{Another important direction is the development of a macroscopic fluctuation theory analogue of the present construction, where the empirical observables entering the rates become spatially extended fields rather than finite-dimensional quantities.} The variational structure of the level-2.5 rate function naturally connects with stochastic control formulations, suggesting alternative approaches to the characterisation of rare events. From a physical standpoint, the results contribute to ongoing efforts to generalise stochastic thermodynamics beyond Markovian dynamics, complementing existing work on semi-Markov processes, delayed systems, and coarse-grained descriptions. \newtext{It would also be interesting to extend the definition of the memory observable by introducing kernels that damp the contribution of the distant past, reset memory, or restrict it to a finite time window. Such extensions would be relevant, for example, for biophysical systems in which the memorised observable is an environmental trace with finite lifetime, such as evaporating chemical trails in autochemotactic motion.} More broadly, SIJPs provide a controlled setting to investigate how memory, feedback, and information encoded in empirical observables shape fluctuations, irreversibility, and precision--dissipation trade-offs, with potential relevance for adaptive biological systems, active matter, and learning-driven stochastic processes.

\acknowledgments

The authors warmly thank Amarjit Budhiraja for many insightful and stimulating discussions. F.C.\ is grateful to Tobias Grafke for help on clarifying numerical aspects behind the example in Section \ref{subsec:twostate}. F.C.\ is supported by a Leverhulme Early Career Fellowship (ECF-2025-482). J.P.G.\ is supported by EPSRC Grants No.\ EP/V031201/1 and EP/T022140/1. 

\appendix

\newtext{\section{Details on SIJP assumptions and counterexamples}
\label{app:ass}

In this Appendix, we clarify the role of the fixed-point argument. In the simpler case $g_{xy}=0$, the observable is bounded by construction and any limiting value of $A_t$ belongs to the compact interval
\begin{equation}
    \mathcal{I}_f
    =
    \left[
    \min_{x\in\mathcal S} f_x,
    \max_{x\in\mathcal S} f_x
    \right].
\end{equation}
The frozen average
\begin{equation}
    \Psi_f(a)
    =
    \sum_{x\in\mathcal S} f_x \gamma_x(a)
\end{equation}
is continuous and maps $\mathcal{I}_f$ into itself. Hence Brouwer's fixed-point theorem guarantees the existence of at least one $a^\star\in\mathcal I_f$ such that
\begin{equation}
    a^\star=\Psi_f(a^\star),
    \qquad
    \gamma(a^\star)Q(a^\star)=0.
\end{equation}
This proves the existence of a self-consistent frozen stationary point, but not convergence of the SIJP towards it. Convergence requires additional stability assumptions on the limiting dynamics. In the scalar case, this corresponds to stability of the fixed points of the limiting ODE
\begin{equation}
    \dot a = \Psi_f(a)-a .
\end{equation}

When $g\neq 0$, boundedness of $f$ and $g$ alone is not sufficient to produce a compact set for $A_t$. We further require that the empirical jump fluxes $\Phi_{xy}(t)$ are bounded and this can be obtained by imposing a uniform bound on the jump rates on the relevant range of $a$. Under such an assumption, one can formulate the self-consistency problem for stationary pairs as
\begin{align}
    \bar a
    &=
    \sum_{x\in\mathcal S} f_x \pi_x
    +
    \sum_{(x,y)\in\E} g_{xy}\varphi_{xy} \\
    \pi Q(\bar{a})&=0 \\
    \sum_{y\neq x}\varphi_{xy}
    &=
    \sum_{y\neq x}\varphi_{yx} \, .
\end{align}
The existence of such a pair is then part of the standing assumptions defining $\mathcal A_{\mathrm{stat}}$.

There are simple ways in which these assumptions can fail. First, imagine a feedback that dynamically removes part of the state space. Take $\mathcal S=\{0,1\}$, $g=0$, $f_0=0$, $f_1=1$, so that
\begin{equation}
    A_t=L_1(t)\in[0,1].
\end{equation}
Then, let
\begin{equation}
    Q_{01}(a)=
    \begin{cases}
        1, & a\geq 1/2,\\
        0, & a<1/2,
    \end{cases}
    \qquad
    Q_{10}(a)=1 .
\end{equation}
This is a discontinuous map that induces a breaking of irreducibility: when $A_t\geq 1/2$, both states communicate and the process is irreducible, however, as soon as $A_t<1/2$, the transition $0\to1$ is switched off. State $0$ then becomes absorbing, and the effective state space collapses from $\{0,1\}$ to the single stae $\{0\}$. Consequently, once the trajectory enters this regime and reaches state $0$, it remains there, and
\begin{align}
    L_0(t)\to 1
    &\qquad
    L_1(t)\to 0 \\ 
    A_t \to 0
    &\qquad
    \Phi_{01}(t),\Phi_{10}(t)\to 0 \, .
\end{align}
Physically, this corresponds to a memory-induced collapse into an inactive state. The long-time behaviour is therefore selected by a reducible limiting dynamics, not by an irreducible frozen generator.

Another interesting example not covered by our theory is obtained from a specific self-exciting activity feedback. Take $\mathcal S=\{0,1\}$, $f_0=f_1=0$, $g_{01}=g_{10}=1$, so that
\begin{equation}
    A_t=\Phi_{01}(t)+\Phi_{10}(t)=\frac{N_t}{t} \, ,
\end{equation}
where $N_t$ is the total number of jumps up to time $t$. Let
\begin{equation}
    Q_{01}(a)=Q_{10}(a)=1+a,
    \qquad a\geq 0 \, .
\end{equation}
For every finite $a$, the frozen chain is irreducible. However, independently of the current state of the process, the instantaneous escape rate is $1+A_t$. Since $A_t=N_t/t$, its deterministic drift is
\begin{equation}
    \frac{d A_t}{dt}
    =
    \frac{1}{t} \, .
\end{equation}
Thus $A_t$ grows as $\log t$ rather than converging to a finite stationary value. Therefore no stationary limiting pair $(\pi,\varphi)$ of the type assumed above exists.

\section{Control of the error due to time-scale separation and block-averaging calculation}
\label{app:err}

We detail here why the error term in Eq.\ \eqref{eq:25Step3} is subextensive in the long-time limit. We denote
\begin{equation}
\begin{split}
F(t) &= \sum_{(x,y)\in\E} \bigg[ \tilde{\rho}_x(t) \big( Q_{xy}(A_t)-\tilde{H}_{xy}(t) \big) \\
&\hspace{0.5cm}+ \tilde{\rho}_x(t)\tilde{H}_{xy}(t) \log \frac{\tilde{H}_{xy}(t)}{Q_{xy}(A_t)} \bigg] \, .
\end{split}
\end{equation}

The error $\text{Err}_T(\Delta t)$ has two contributions. The first comes from freezing the time-dependent cost density $F(t)$ over each block, viz.\ it comes from replacing an integral with a sum. The second comes from replacing empirical occupations and fluxes within a block by the frozen accompanying distribution.

We first consider the freezing error. On the block $[t_k,t_k+\Delta t]$, this error is
\begin{equation}
\begin{split}
E_k^{\text{freeze}} &= \int_{t_k}^{t_k+\Delta t} F(s)\,ds - F(t_k)\Delta t \\
&=
\int_{t_k}^{t_k+\Delta t} \big[ F(s)-F(t_k) \big]\,ds \, .
\end{split}
\end{equation}
The time-scale separation assumption is equivalently expressed by requiring the frozen cost density to vary on the slow SIJP time scale,
\begin{equation}
\left| \frac{dF}{dt}(t) \right| = O(t^{-1}) \, .
\end{equation}
Therefore, for $s \in [t_k,t_k+\Delta t]$,
\begin{equation}
\begin{split}
\big| F(s)-F(t_k) \big| &\leq C \int_{t_k}^{s} \frac{du}{u} = C \log\frac{s}{t_k} \\
&= C \log \left( 1+\frac{s-t_k}{t_k} \right) \, .
\end{split}
\end{equation}
Since $\Delta t \ll t_k$ on the slow scale, this implies
\begin{equation}
\log \left( 1+\frac{s-t_k}{t_k} \right) \leq \frac{s-t_k}{t_k} \, .
\end{equation}
Hence
\begin{equation}
\big| E_k^{\text{freeze}} \big| \leq \frac{C}{t_k} \int_{t_k}^{t_k+\Delta t} (s-t_k)\,ds = O\left( \frac{(\Delta t)^2}{t_k} \right) \, .
\end{equation}
Summing over blocks gives
\begin{equation}
\sum_k \big| E_k^{\text{freeze}} \big| = O\left( \sum_k \frac{(\Delta t)^2}{t_k} \right) \, .
\end{equation}
Using $t_k=k\Delta t$, up to an initial finite-time block which gives only a subextensive contribution,
\begin{equation}
\sum_k \frac{(\Delta t)^2}{t_k} = \Delta t \sum_k \frac{1}{k} =
O(\Delta t\log T) \, .
\end{equation}
Thus
\begin{equation}
E_T^{\text{freeze}} = O(\Delta t\log T)\, .
\end{equation}

We now consider the second source of error. Within each block, the empirical occupations and fluxes are replaced by the frozen accompanying distribution, as in Eqs.\ \eqref{eq:FrozenAccompany1} and \eqref{eq:FrozenAccompany2}. This replacement is valid after the frozen dynamics has relaxed. If $\tau_{\text{micro}}$ denotes a uniform upper bound on the mixing times of the frozen generators, the non-stationary transient contributes an error of order $\tau_{\text{micro}}$ per block. Since the number of blocks is of order $T/\Delta t$, the total mixing error is
\begin{equation}
E_T^{\text{mix}}
=
O\left(
\frac{T}{\Delta t}
\tau_{\text{micro}}
\right) \, .
\end{equation}

Combining the two estimates, we obtain
\begin{equation}
\frac{|\text{Err}_T(\Delta t)|}{T} = O\left( \frac{\Delta t\log T}{T} + \frac{\tau_{\text{micro}}}{\Delta t} \right) \, .
\end{equation}
Therefore, choosing for instance $\Delta t=T^\alpha$ with $0<\alpha<1$, while keeping $\tau_{\text{micro}}=O(1)$, gives
\begin{equation}
\frac{|\text{Err}_T(\Delta t)|}{T} \to 0 \, ,
\end{equation}
as $T\to\infty$.
Thus the accumulated block error is subextensive and does not contribute to the large-deviation rate function.}

\section{Simulating a SIJP}
\label{app:sim}

We now detail how to run simulations of SIJPs. Naively applying a \deletetext{Gillepsie}\newtext{Gillespie} algorithm \deletetext{won't}\newtext{does not} work as the rate matrix $Q(A_t)$ of a SIJP changes with time due to the fact that the empirical observable \eqref{eq:EmpDepSIJP} is explicitly time-dependent.

The survival probability, viz.\ the probability to be at, say, state $x \in \mathcal{S}$ for a time longer than $u$, having landed onto $x$ at time $t$\newtext{,} is therefore a time-dependent quantity and takes the form
\begin{equation}
\label{eq:SurvivalSIJP}
    \exp \left( \int_t^{t+u} Q_{x x}(A_{t'}) \, dt' \right) \, ,
\end{equation}
where the diagonal term $Q_{xx}(A_{t'})$ is defined in \er{eq:DiagRateSIJP}. \deletetext{Thus, we can write the time-dependent mean escape rate $\Lambda_{x}(t,t+u)$ (whose inverse is the mean waiting time) to hop out of $x$ as in \er{eq:JumpOutRateSIJP}.}\newtext{Equivalently, introducing the instantaneous escape rate
\begin{equation}
    \lambda_x(t') \coloneqq - Q_{xx}(A_{t'})
    =
    \sum_{y\neq x} Q_{xy}(A_{t'}) \, ,
\end{equation}
we define the corresponding integrated escape rate as
\begin{equation}
    \Lambda_x(t,t+u)
    \coloneqq
    \int_t^{t+u} \lambda_x(t')\,dt'
    =
    -\int_t^{t+u} Q_{xx}(A_{t'})\,dt' \, ,
\end{equation}
as in \er{eq:JumpOutRateSIJP}.}

A trajectory of \deletetext{lenght}\newtext{length} $T$ for such a SIJP can then be realised in two equivalent ways. First, we can implement an accept-reject algorithm, which is based on the (Ogata) thinning construction, see also Ex.\ 2.1.6 in~\cite{daley2003an-introduction}. Imagine the process sits at state $x$ at time $t$. \deletetext{The mean escape rate $\Lambda_{x}(t,t+u)$ for all $u \geq 0$ in the time horizon of interest can be bounded by $\bar{\Lambda}_x \geq \Lambda_{x}(t,t+u)$.}\newtext{The instantaneous escape rate can be bounded on the time horizon of interest by a constant $\bar{\lambda}_x$ such that
\begin{equation}
    \bar{\lambda}_x \geq \lambda_x(t')
    =
    -Q_{xx}(A_{t'})
\end{equation}
for all relevant $t'\in[t,T]$. Such a bound can be chosen, for instance, from an a priori bound on the admissible values of $A_{t'}$, or updated adaptively after each rejected proposal.} Then, it is enough to (i) propose exponential waiting times $u$ with rate \deletetext{$\bar{\Lambda}_x$}\newtext{$\bar{\lambda}_x$} and (ii) accept them with probability \deletetext{$\Lambda_{x}(t,t+u)/\bar{\Lambda}_x(t,t+u)$}\newtext{$\lambda_x(t+u)/\bar{\lambda}_x$}. If accepted, (iii) we update time, empirical observable, and choose next state $y \in \mathcal{S}$ with probability
    \begin{equation}
        \label{eq:ProbaNextState}
        \frac{Q_{xy}(A_{t+u})}{\deletetext{\Lambda_{x}(t,t+u)}\newtext{\lambda_x(t+u)}} \, .
    \end{equation}
Otherwise, (iii) we shift time by $u$, i.e., $t \leftarrow t+u$, keep the process at $x$, update the empirical observable, compute a new bound \deletetext{$\bar{\Lambda}_x$}\newtext{$\bar{\lambda}_x$} and loop back to (i), repeating until $t+u$ exceeds the final observation time $T$.

A second method that can be used to simulate SIJPs is based on direct numerical integration. At time $t$, while the process sits at $x \in \mathcal{S}$, (i) we extract \deletetext{$E \sim \exp(1)$}\newtext{$E \sim \mathrm{Exp}(1)$, namely a unit-rate exponential random variable}, and (ii) numerically solve $\Lambda_{x}(t,t+u) = E$ for $u \geq 0$ using, e.g., bisection methods. Then, (iii) we advance to time $t \leftarrow t+u$, update the empirical observable, and choose next state $y \in \mathcal{S}$ with probability \eqref{eq:ProbaNextState}. We loop back to (i) and iterate until $t+u$ exceeds $T$.

The first method is cheaper in terms of computational resources, but it might be affected by a low acceptance rate, which would increase the overall simulation time. The second method is heavier than the first from the computational point of view as quadrature techniques need to be included. However, it is not affected by an accept-reject step. In all the examples discussed in this work we made use of the second algorithm.

We conclude by remarking that the construction above and the algorithms mentioned are not restricted to SIJPs. They can be applied to---in fact, they were developed to work with---time-inhomogeneous Markov processes. In our SIJP, the time non-homogeneity is implicit and is evaluated through the empirical observable. We refer the reader to Chap.\ 2 of \cite{daley2003an-introduction} for further details on how to treat time-inhomogeneous Poisson processes.

\bibliography{bibliography-19032026}

\end{document}